\documentclass[twocolumn,prd,superscriptaddress,showpacs,amsmath,amssymb]{revtex4-1}

\usepackage{graphicx}
\usepackage{dcolumn}
\usepackage{bm}
\usepackage{multirow}

\begin{document}

\title{CoGeNT: A Search for Low-Mass Dark Matter using $p$-type Point Contact Germanium Detectors}

\def\PNNL{Pacific Northwest Laboratory, Richland, WA 99352, USA}
\def\UC{Kavli Institute for Cosmological Physics and Enrico Fermi Institute, University of Chicago, Chicago, IL 60637, USA}
\def\CANBERRA{CANBERRA Industries, Meriden, CT 06450, USA}
\def\UW{Center for Experimental Nuclear Physics and Astrophysics and Department of Physics, University of Washington, Seattle, WA 98195, USA}

\author{C.E.~Aalseth} \affiliation{\PNNL}
\author{P.S.~Barbeau} \altaffiliation{Present address: Department of Physics, Stanford University, Stanford, CA 94305, USA} \affiliation{\UC} 
\author{J.~Colaresi} \affiliation{\CANBERRA}
\author{J.I.~Collar} \email{Electronic address: collar@uchicago.edu}\affiliation{\UC}
\author{J.~Diaz Leon} \affiliation{\UW}
\author{J.E.~Fast} \affiliation{\PNNL}
\author{N.E.~Fields} \affiliation{\UC}
\author{T.W.~Hossbach} \affiliation{\PNNL} \affiliation{\UC}
\author{A.~Knecht} \affiliation{\UW}
\author{M.S.~Kos} \email{Electronic address: marek.kos@pnnl.gov}\affiliation{\PNNL}
\author{M.G.~Marino} \altaffiliation{Present address: Physics Department, Technische Universit\"at M\"unchen, Munich, Germany}\affiliation{\UW}
\author{H.S.~Miley} \affiliation{\PNNL}
\author{M.L.~Miller} \altaffiliation{Present address: Cloudant - West Coast, 209 1/2 1$^{st}$ Ave S,  Seattle, WA 98104}\affiliation{\UW}
\author{J.L.~Orrell}  \affiliation{\PNNL}
\author{K.M.~Yocum} \affiliation{\CANBERRA}

\collaboration{CoGeNT Collaboration}

\date{\today}

\begin{abstract}
CoGeNT employs p-type point-contact (PPC) germanium detectors to search for Weakly Interacting Massive Particles (WIMPs). By virtue of its low energy threshold and ability to reject surface backgrounds, this type of device allows an emphasis on low-mass dark matter candidates ($m_{\chi}\sim10$~GeV/c$^{2}$).  We report on the characteristics of the PPC detector presently taking data at the Soudan Underground Laboratory, elaborating on aspects of shielding, data acquisition, instrumental stability, data analysis, and background estimation.  A detailed background model is used to investigate the low energy excess of events previously reported, and to assess the possibility of temporal modulations in the low-energy event rate.  Extensive simulations of all presently known backgrounds do not provide a viable background explanation for the excess of low-energy events in the CoGeNT data, or the previously observed temporal variation in the event rate.  Also reported on for the first time is a determination of the surface (slow pulse rise time) event contamination in the data as a function of energy.  We conclude that the CoGeNT detector technology is well suited to search for the annual modulation signature expected from dark matter particle interactions in the region of WIMP mass and coupling favored by the DAMA/LIBRA results.
\end{abstract}



\maketitle

\section{Introduction}

CoGeNT (Coherent Germanium Neutrino Technology) is a program aiming to exploit the characteristics of p-type point-contact germanium detectors in areas as diverse as the search for low-mass dark matter candidates, coherent neutrino-nucleus elastic scattering, and $^{76}$Ge double-beta decay \cite{jcap}. 

Data collected from a first CoGeNT detector at a shallow underground location demonstrated sensitivity to low-mass ($<10$~GeV/c$^{2}$) dark matter particles \cite{Aal08}. In particular, it appeared CoGeNT was particularly well suited to address the DAMA/LIBRA \cite{DAMA} modulation result. Following the identification of several sources of internal background in this prototype, a second CoGeNT detector was installed in the Soudan Underground Laboratory (SUL) during 2009 with the goal of improving upon the dark matter sensitivity reach of the 2008 result \cite{Aal08}. The first 56-days of operation of the CoGeNT detector at SUL showed an unexpected excess of events \cite{Aal11} above the anticipated backgrounds for ionization energies below 2~keV. Further data collection from this detector continued until an interruption imposed by a fire in the access shaft to the laboratory halted the initial run in March of 2011. Analysis of the accumulated data set \cite{Aal11b}, spanning 442 live days over the period 4 December 2009 to 6 March 2011, showed a $\sim2.8\sigma$ significance modulation of the monthly event rate in the low-energy region that is compatible with the dark matter signature described in \cite{andrzej}. The fitting procedure generating this low-significance modulation result used unconstrained phase, period, and amplitude variables. Time-stamped data have been made publicly available, allowing for a number of independent analyses and interpretations.

In this paper we provide a more in-depth description of the apparatus and data analysis, concentrating on aspects of instrument stability, data cuts, uncertainties, and background estimation. The data set employed for this discussion is the same as in \cite{Aal11b}, and all energies are in keVee (keV electron equivalent, i.e., ionization energy), unless otherwise stated. Following the three-month outage resulting from the Soudan fire, this detector has taken data continuously, starting 7 June 2011. An additional body of data is to be released in the near future. The design and expectations for CoGeNT-4 (C-4), a planned expansion aiming at an increase in active mass by a factor of ten, featuring four large PPC detectors with a reduced energy threshold and lower background, are discussed in a separate publication \cite{inprep}.

\section{Description of the Apparatus}

The present CoGeNT detector is located at the Soudan Underground Laboratory (Soudan, Minnesota, USA) at a vertical depth of 2341~feet (689~feet below sea level), providing 2090~meters of water equivalent (m.w.e.) overburden as shielding against cosmic rays and associated backgrounds. The detector shield is placed on a floor built on top of base I-beams that once supported the Soudan-2 proton decay experiment \cite{All96}. The detector element is a single modified BEGe germanium diode. BEGe (Broad Energy Germanium) is the commercial denomination used by the manufacturer (CANBERRA Industries) for their line of PPC detectors. The technical characteristics of this PPC are shown in Table \ref{tab:BEGeCharacteristics}. The detector is contained within an OFHC copper end cap cryostat, and mounted in an OFHC copper inner can connected to an OFHC copper cold finger. Internal detector parts were custom manufactured in either OFHC copper or PTFE. All internal parts were etched to remove surface contaminations using ultra-pure acids in class 100 clean room conditions, following procedures similar to those described in \cite{etch}. A commercial stainless steel horizontal cryostat encloses the rear of the assembly, providing electrical feed-through to a side-mounted CANBERRA DPRP pulse-reset preamplifier typically used in high-resolution X-ray detectors (figure \ref{fig:BEGeInnerShield}).

\begin{center}
\begin{table}[b]
\caption{\label{tab:BEGeCharacteristics} Characteristics of the CoGeNT high purity PPC germanium detector at SUL.}
\begin{tabular}{lr} \hline
Property & Value \\ \hline
Manufacturer & CANBERRA (modified BEGe)\\
Total Mass & 443 gram \\
Estimated Fiducial Mass & $\sim$330 gram \\
Outer Diameter & 60.5 mm \\
Length & 31 mm \\
Capacitance & 1.8 pF (at 3000~V bias) \\ \hline
\end{tabular}
\end{table}
\end{center}

\subsection{Shield design}

\begin{figure}[!htbp]
\includegraphics[height=0.27\textheight]{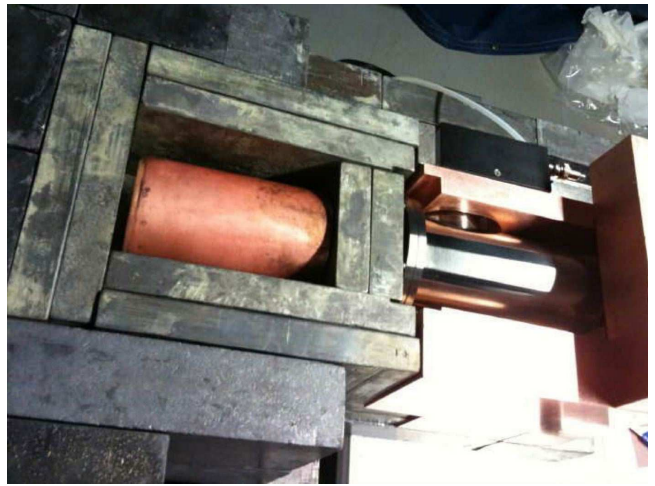}
\caption{\label{fig:BEGeInnerShield} Partially disassembled shield of the CoGeNT detector at SUL, showing the cylindrical OFHC end cap and innermost 5~cm of ancient 0.02~Bq $^{210}$Pb/kg lead, characteristically oxidized following etching. The preamplifier is visible at the top right (black box). A minimum of 7~cm of lead thickness shields the detector from the naturally occuring radioactivity in the preamplifier's electronic components.}
\end{figure} 

The lead shield involves three categories of lead bricks. The innermost 5~cm layer is composed of acid-etched ultra-low background ancient lead having a $^{210}$Pb content of approximately 0.02~Bq $^{210}$Pb/kg, measured using radiochemical extraction followed by alpha spectroscopy at PNNL \cite{SMiley}. This layer provides shielding against the $^{210}$Pb bremsstrahlung continuum from external contemporary lead, resulting in a negligible low-energy background from this source of less than 0.01~counts / keVee / kg-Ge / day \cite{pb210}. OFHC copper bricks are used to provide mechanical support around the stainless steel horizontal cryostat body (figure \ref{fig:BEGeInnerShield}). A middle 10~cm thick layer of contemporary ($\sim$100~Bq $^{210}$Pb/kg) lead bricks is also chemically etched and cleaned. The outer 10~cm thick layer is composed of stock bricks not chemically etched. A minimum of 25~cm of lead surround the detector element in all directions. The assembly of the lead shield was performed inside a temporary soft-wall clean room, to avoid excess dust. 

\begin{figure}[!htbp]
\includegraphics[height=0.35\textheight]{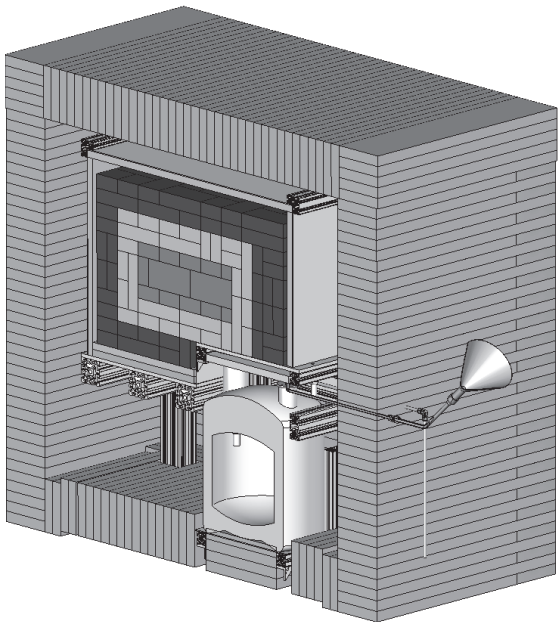}
\caption{\label{fig:SoudanOuterShield} Layout of the complete shield for the CoGeNT detector. The outermost component is a layer of recycled HDPE, used to moderate neutrons.  Next towards the interior, a 1~inch thick layer of borated polyethylene captures moderated neutrons.  Three layers of lead are indicated by the three different inner shaded regions.  The outermost lead is composed of stock bricks, not chemically etched, the middle layer is chemically etched and cleaned, and the innermost layer consists of ultra-low background ancient lead.  An automated liquid nitrogen transfer system refills the detector Dewar every 48~hours, maintaining the germanium crystal at a near constant temperature. See text for a full description of these components.}
\end{figure} 

Exterior to the lead shield is a 2.5~cm thick layer of 30\% borated polyethylene, intended to act as a thermal neutron absorber. The borated polyethylene panels are sealed using heavy vinyl tape as a barrier against radon ingress. The inner lead shield and the borated polyethylene are contained inside of an aluminum sheet-metal box (table base, four walls, and top). All edges are once again sealed using heavy vinyl tape. Shielding materials internal to this radon-exclusion volume are supported by an aluminum extrusion table approximately 66~cm above the floor. This volume is continuously flushed with boil-off nitrogen gas from a dedicated pressurized Dewar, at a rate of 2~liters per minute. An extruded aluminum structural frame provides mechanical rigidity to the sealed aluminum box. The detector Dewar rests on a layer of vibration absorbing foam aiming at reducing microphonic events (Sec.IV). Finally, an external layer of recycled high-density polyethylene (HDPE) deck planking is used to enclose the entire assembly, acting as a neutron moderator. The HPDE is 18.3~cm thick, with nearly complete $4\pi$ coverage (the only breach being the table legs supporting the lead cave). These elements can be seen in figure \ref{fig:SoudanOuterShield}.

Not visible in figure \ref{fig:SoudanOuterShield} is an active muon veto composed of 10 flat panels surrounding the HDPE shield, with six 120~cm $\times$ 120~cm panels on the sides and four 100~cm $\times$ 100~cm panels covering the top with considerable overlap and overhang. The veto panels are 1~cm thick and read-out via a single PMT located at the center of each panel. The light collection efficiency was measured at a grid of positions in the panels using a low-energy gamma source, observing a minimum yield at all locations better than 50\% of the central maximum. A $\sim$90\% geometric coverage of the shield is estimated for this muon veto. Further discussion of its efficiency is provided in Sec.IV-A.

\subsection{Data acquisition}

\begin{widetext}

\begin{figure}[!htbp]
\includegraphics[width=1.\textwidth]{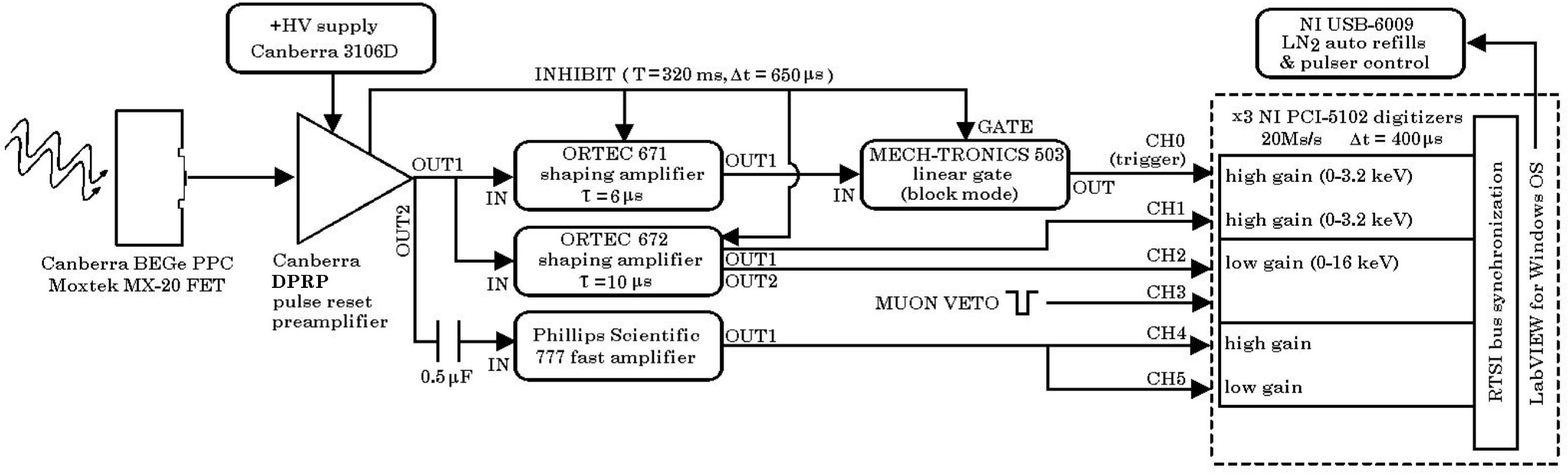}
\caption{\label{fig:electronics} Schematic of the data acquisition system for the CoGeNT detector at SUL (see text). }
\end{figure}

\end{widetext}

Figure \ref{fig:electronics} shows an schematic of the data acquisition (DAQ) system used in the present CoGeNT installation at SUL \cite{phil}. It combines analog amplification of detector pulses with digitization of raw preamplifier traces, the second permitting the rejection of events taking place near the surface of the germanium crystal via rise time cuts \cite{Aal11}. An initial data taking period from the end of August 2009 to 1 December 2009 did not include preamplifier trace digitization. This period allowed for the decay of short-lived cosmogenic isotopes (e.g., $^{71}$Ge with $t_{1/2} = 11.4$~d). In early December 2009 a third National Instruments PCI-5102 digitizer card was installed to collect preamplifier traces. During this initial period a parallel DAQ system based on the GRETINA Mark IV digitizer \cite{gretina} was also tested, but found to provide limited information for low energy analysis \cite{mike}.

A pulse-reset preamplifier, typically employed for silicon X-ray detectors, is used in combination with a field-effect transistor (FET) specially selected to match the PPC's small ($\sim$2 pF) capacitance. This allows for the lowest possible electronic noise and energy threshold \cite{jcap}. The preamplifier generates two equivalent signal outputs, an inhibit logic signal when the pulse reset circuitry of the preamplifier is active, and accepts a test input (electronic pulser). The test input is normally disconnected, terminated, and isolated to avoid spurious noise injections. While the ORTEC 671 and 672 shaping amplifiers utilize the inhibit logic signal to protect against distortions caused by the preamplifier reset, the amplifier outputs are sufficiently altered to initiate the DAQ, which is set to trigger on very low energy (300~eVee) shaped pulses. Even with the very long reset period ($\sim$320~ms) achieved in this detector --- a result of its sub-pA leakage current --- this would generate an unacceptable $\sim300$~Gbyte/day of pulse reset induced traces streaming to disk. The triggering output of the 671 shaping amplifier is therefore further inhibited by use of a linear gate operated in blocking mode. The gate is observed to add a negligible amount of noise to the already sufficiently amplified pulses. The duration of the inhibit logic pulse is set to its maximum (650~$\mu$s) in order to ensure a complete restoration of the amplifier baseline following resets (achieved within $\sim100$~$\mu$s), while generating a negligible 0.2\% dead time. The frequency of the preamplifier resets, which is directly proportional to the leakage current of the detector and in turn to the germanium crystal temperature, has been periodically measured and shown to have remained constant thus far. Any significant alteration of this leakage current would also appear as a measurable increase in the white parallel component of the detector noise \cite{pullia}, dominant for the channel used in noise monitoring (shaping time $\tau=10$~$\mu$s). The detector noise is observed to be very stable over the detector's operational period (figure~\ref{fig:stability}). Further discussion on DAQ stability is provided in Sec.III-E. 

\begin{figure}[!htbp]
\includegraphics[width=0.46\textwidth]{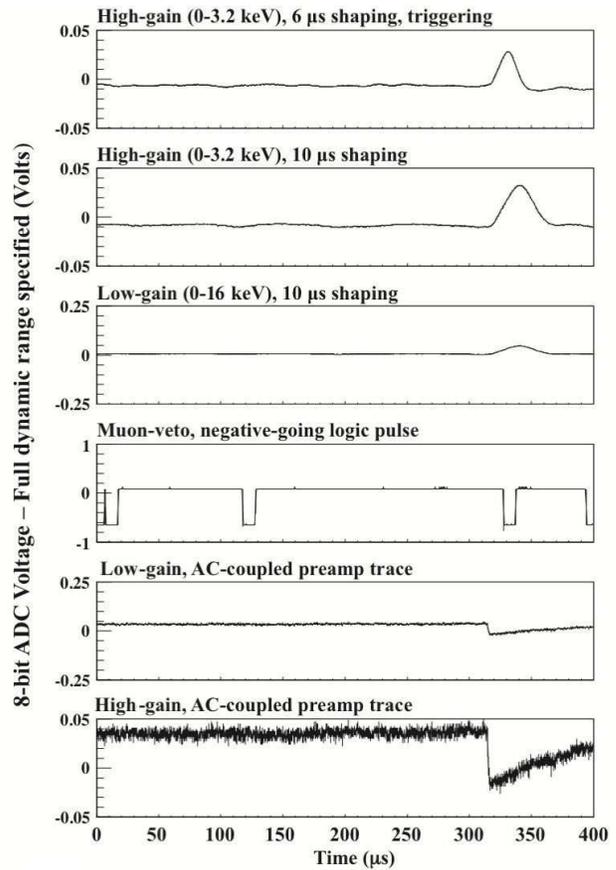}
\caption{\label{fig:CoGeNT-Traces} Example digitized traces from the six CoGeNT DAQ read-out channels, corresponding to an event with energy $\sim2.5$~keVee. Preamplifier traces are DC-offset at the Phillips Scientific 777 amplifier to allow for rise time measurements of pulses in the range 0-12~keVee, following offline wavelet denoising \protect\cite{Aal11} (not yet applied to these traces).}
\end{figure}

The readout system is composed of three hardware-synchronized PCI-based National Instruments digitizers totalling 6 channels, sampling at 20~MSamples/s, each with a resolution of 8~bits. The acquisition software is a Windows-based LabVIEW program, also responsible for liquid nitrogen auto-refills and electronic pulser control. Raw preamplifier traces are amplified prior to digitization using a low-noise Phillips Scientific 777 fast amplifier (200~MHz bandwidth), using a DC-blocking capacitor at its input to yield a $\sim$50~$\mu$s preamplifier pulse decay time, noticeable in figure~\ref{fig:CoGeNT-Traces}. 

Following gain-matching bias adjustments, the PMT outputs from all muon veto panels are daisy-chained and reduced to one single channel, which is linearly amplified, discriminated with a threshold set at single photo-electron level, and further conditioned using a gate generator, the output of which is digitized by the DAQ \cite{phil}. Traces captured for an example event are shown in Figure~\ref{fig:CoGeNT-Traces}. Digitized trace lengths are an intentionally long at 400~$\mu$s, with 80\% pre-trigger content. Pre-trigger information allows for pulse diagnostics (Sec.IV), monitoring of detector noise and trigger threshold stability (Sec.III-E), and is also used in pulse simulations (Sec.IV-B).

The PC housing the digitizer cards maintains an internal buffer to store a set of events. After 20~events are stored, data from the digitizer buffer is written to disk. File names are cycled (open file closed, saved, and new file opened) every 3~hours. Data are automatically backed-up to a second PC, from which they are transferred to a remote server. 

\section{Detector Characterization}

Several aspects of detector and DAQ characterization are described in this section. 

\subsection{Energy Calibration}

The existing DAQ system was developed with an emphasis on instrumental stability, minimization of electronic noise, and on providing a maximum of information about low-energy events. It is however limited in its energy range, 0-16~keVee. While it is possible to increase this range during dedicated background characterization runs (figure~\ref{fig:data300gammas}), this can be done only at the expense of valuable information used for data selection cuts at lower energies. During normal operation, no viable external gamma sources exist for low-energy calibration. This is due to the thickness of the OFHC cryostat parts and germanium dead layer surrounding the active bulk of the detector, which dramatically attenuate external low-energy photons. Fortunately, a number of internal peaks arising from cosmogenic isotopes decaying via electron capture (EC) are visible in the region 1-10~keVee. These are used to extract an accurate energy calibration and to characterize the energy resolution as a function of energy. The reader is referred to \cite{Aal08,Aal11,Aal11b} for additional details. 

\subsection{Quenching Factor}

\begin{figure}[!htbp]
\includegraphics[width=0.46\textwidth]{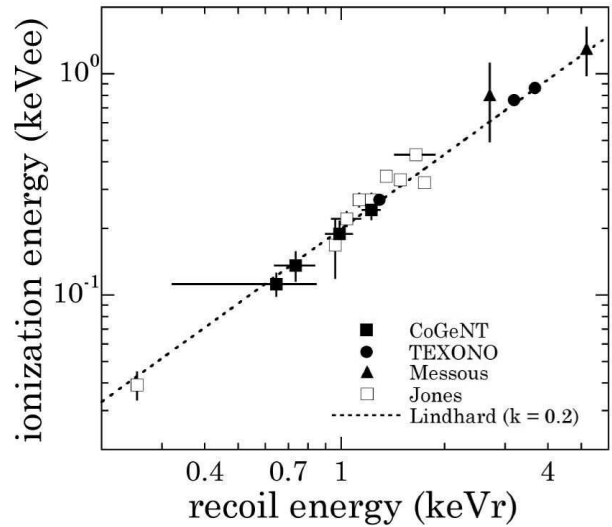}
\caption{\label{fig:quenching} Neutron scattering measurements of the low-energy quenching factor for nuclear recoils in germanium, compared to Lindhard theory predictions.  CoGeNT adopts the expression relating ionization and recoil energy $E_{i}$(keVee)~$= 0.2 \times E_{r}^{1.12}$(keVr), valid for the range 0.2~keVr~$<E_{r}<$~10~keVr, and essentially indistinguishable from the Lindhard case plotted.}
\end{figure}

The quenching factor, defined as the measurable fraction of the energy deposited by a nuclear recoil in a detecting medium, is a quantity of particular relevance for WIMP dark matter studies. For PPCs and conventional germanium detectors, its characterization involves a measurement of the ionization generated by a discrete recoil energy, typically induced in a neutron calibration. The CoGeNT PPC described in \cite{Aal08} was exposed to a custom-built monochromatic 24~keV filtered neutron beam at the Kansas State University research reactor. This PPC crystal is nearly identical to that operating in SUL \cite{Aal11,Aal11b} (BEGe contact geometry, similar 160~eVee FWHM electronic noise and 0.5~keVee threshold, 83.4~cc {\frenchspacing vs.} 85~cc crystal volume, and the same nominal Li diffusion depth in the outer contact). Triggering on the neutron capture peak of the $^{6}$LiI scintillator used to detect the scattered neutrons \cite{phil} allowed the measurement of sub-keV quenching factors, found to be in good agreement with other available data (figure~\ref{fig:quenching}). Details on neutron beam design and characterization, and on the analysis of these data are provided in \cite{reactor} and \cite{phil}, respectively.

\subsection{Dead Layer}

\begin{figure}[!htbp]
\includegraphics[width=0.46\textwidth]{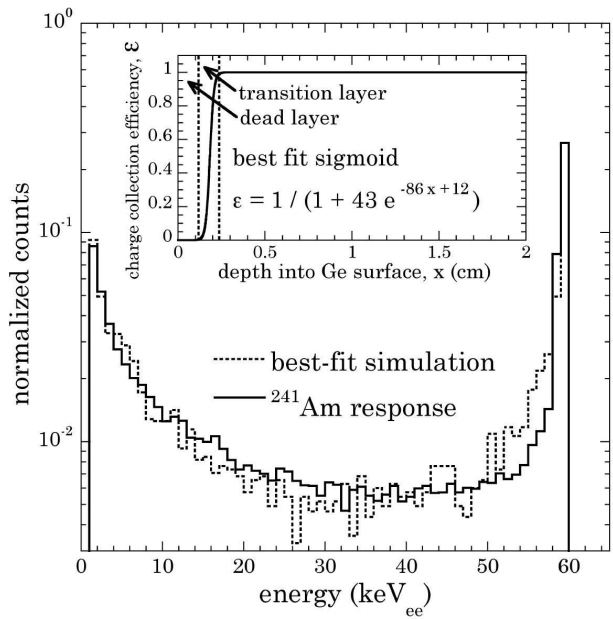}
\caption{\label{fig:deadlayer} Characterization of surface structure on the external n+ contact of a PPC (see text). The two free sigmoid parameters are fit via comparison of calibration data and Monte Carlo simulation. Energy depositions taking place in the transition layer near its boundary with the dead layer lead to large signal rise times, i.e., slow pulses. On the opposite side of the transition layer, rise times progressively approach the small values typical of a fast (bulk) event.}
\end{figure}

PPC detectors feature an inert outer contact layer over most of their surface. The depth of this dead layer can be tuned during the manufacturing process, by controlling the amount of lithium diffused into this region. CoGeNT detectors are built with the maximum diffusion depth possible during BEGe fabrication, nominally a $\sim1$ mm dead layer over all surfaces except for a small (3.8 cm$^{2}$) intra-contact passivated area. This dead layer acts as a passive barrier against external low-energy radiation (X-rays, betas, etc.). Events taking place in the region immediately below this dead layer (``transition layer,'' figure~\ref{fig:deadlayer}) generate pulses with a characteristically slow rise time, and a partial charge collection efficiency \cite{Aal11,sakai,ryan}. The surface structure of the CoGeNT PPC in \cite{Aal08} was characterized using uncollimated $^{241}$Am 59.5~keV gammas impinging on the top surface of the germanium crystal, opposite to the central contact. Following a MCNP-Polimi simulation \cite{polimi} of interaction depth {\frenchspacing vs.}  energy deposition including all internal cryostat parts, and assuming a sigmoid description of charge collection efficiency as a function of depth into the crystal, we find a best-fit profile quantitatively and qualitatively similar to that described in \cite{sakai} ($\sim\!1$~mm dead layer, $\sim\!1$~mm transition layer, figure~\ref{fig:deadlayer} inset). This characterization was unfortunately not possible for the PPC at SUL \cite{Aal11, Aal11b} prior to installation within its shield. Due to the aforementioned very similar characteristics for these two PPCs, we adopt the same surface structure when calculating the fiducial (bulk) volume following rise time cuts \cite{Aal11}, while cautiously assigning a $\sim$10\% uncertainty to its value. Additional tests are planned following removal of the PPC at SUL from its shielding.

While the passive shielding provided by the deepest possible lithium diffusion is useful for low-energy background reduction in a dark matter search, it is clearly detrimental to the fiducial mass of a relatively small PPC crystal (Table \ref{tab:BEGeCharacteristics}). This fiducial mass loss due to deep lithium diffusion for background reduction creates a contrast to the requirements of $^{76}$Ge double-beta decay experiments like \textsc{Majorana} \cite{majorana} and GERDA \cite{gerda}, where a maximization of the active enriched germanium mass is preferable. Surface characterization studies using a PPC featuring a shallower lithium diffusion can be found in \cite{ryan} and support the notional model of energy depositions in the transition layer resulting in pulses of partial charge collection and slowed rise times.

\subsection{Trigger Efficiency}

\begin{figure}[!htbp]
\includegraphics[width=0.46\textwidth]{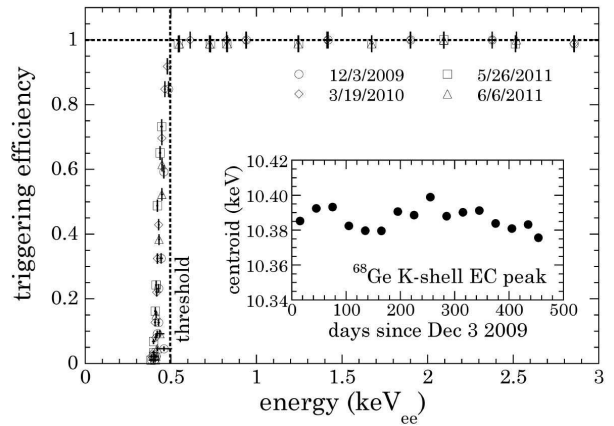}
\caption{\label{fig:triggering} Trigger efficiency vs. energy equivalent for 10~Hz tailed electronic pulses generated with a 814FP CANBERRA pulser. Inset: gain shift stability monitored through the centroid of a Gaussian fit to the 10.3~keV cosmogenic peak. The count rate under this peak decayed from roughly 500 to 150 events per month over the time span plotted.}
\end{figure}

The PPC detector in \cite{Aal11, Aal11b} and its DAQ were operated for a year at a depth of 30~m.w.e., up to a few weeks before installation at SUL. During that time (and the cosmogenic activation ``cooling'' period August-December 2009 at SUL) automatic pulser calibrations were performed for a minute every two hours, revealing an excellent trigger rate stability (better than 0.1\%) for electronic pulses with energy equivalent to 1.85~keVee \cite{phil}. To avoid the injection of any noise or spurious pulses through the preamplifier test input during dark matter search runs, these automatic calibrations were suspended in December of 2009, isolating and terminating that input. However, trigger efficiency calibrations using an electronic pulser have been performed thus far four times, during each interruption to physics runs, yielding reproducible results (figure~\ref{fig:triggering}). These calibrations allow us to calculate triggering efficiency corrections to the energy 
spectrum near threshold, as well as to determine the energy-dependent signal acceptance for fast rise time pulses,  representative of ionization events occurring in the bulk of the crystal \cite{Aal11, Aal11b}. In addition to these pulser calibrations, the trigger threshold level is monitored continuously, as described in the following section.

\subsection{Overall Stability}

\begin{figure}[!htbp]
\includegraphics[width=0.46\textwidth]{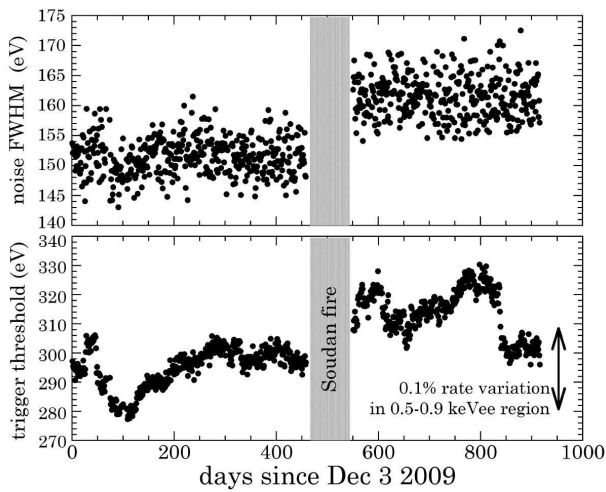}
\caption{\label{fig:stability} Daily average electronic noise and trigger threshold in the CoGeNT PPC at SUL. The small jump in electronic noise post-fire has a negligible effect on the detector threshold. It is the result of either temperature cycling of the crystal (leading to known processes capable of altering the detector leakage current, minimally in this case) or a displacement of cables during emergency post-fire interventions. The fluctuations in trigger threshold agree well with expectations based on manufacturer specifications for the ORTEC 672 shaping amplifier and NI PCI-5102 digitizers, and the observed $\pm1$~$^\circ$C environmental temperature changes measured at SUL.}
\end{figure}

No significant changes in gain have been observed for the PPC at SUL over more than two years of continuous operation, as monitored by the position of the 10.37~keV $^{68}$Ge decay peak (inset figure~\ref{fig:triggering}) and of the energy threshold, immutable at 0.5~keVee.  The long (320~$\mu$s) pre-trigger segment of the traces collected by the DAQ allows us to monitor both the electronic noise of the detector and the small fluctuations in trigger threshold level induced by fluctuations of the CH0 baseline with respect to the constant (i.e., digitally-set) threshold level (figure~\ref{fig:stability}). These baseline fluctuations do not result in a smearing of the energy resolution, given that the zero-energy level is recomputed for each individual pulse from its pre-trigger baseline. They result instead in small shifts by a maximum of $\pm$20~eVee in the sigmoid-like threshold efficiency curve in figure~\ref{fig:triggering}. As a result, they produce correlated changes in trigger rate below the 0.5~keVee threshold, but their effect is negligible above $\sim$0.55~keVee, an energy for which the triggering efficiency reaches 100\%. It is possible to calculate the effect of these baseline fluctuations on the counting rate above the analysis threshold for an exponentially decreasing spectrum like that observed \cite{Aal11b}: this is $\pm$0.1\% for the region 0.5-0.9~keVee (figure~\ref{fig:stability}), and smaller for any energy range extending beyond 0.9~keVee, which is negligible from the point of view of a search for a few percent annual modulation.

\begin{figure}[!htbp]
\includegraphics[width=0.5\textwidth]{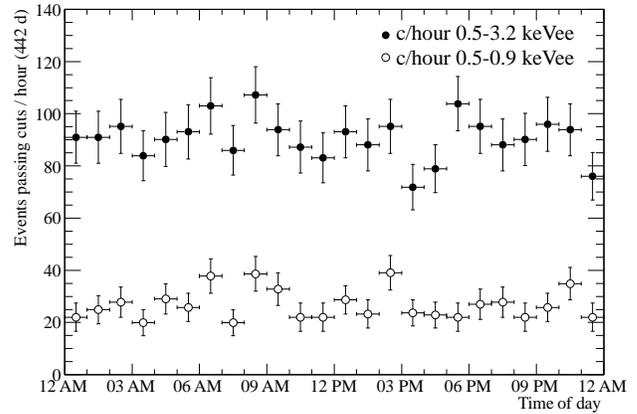}
\caption{\label{fig:diurnal} Diurnal stability of CoGeNT at SUL. Periods of human presence at SUL are $\sim$7 am - 5 pm.}
\end{figure}

Much interest has been traditionally placed on investigating modulated backgrounds having an origin in natural radioactivity (underground muons, radon emanations, etc., see Sec.V), but little discussion can be found in the literature on the specific details of possible instrumental instabilities affecting the DAMA/LIBRA experiment. Searches for a dark matter annual modulation signature need to be concerned about these, in view of the small (few percent) fluctuations in rate expected, the low energies involved, and the unfortunate seasonally correlated phase, having a maximum in summer and minimum in winter, similar to so many unrelated natural processes. As mentioned, it is possible to exclude gain shifts, variations in detector noise and threshold position, and trigger threshold level  fluctuations as sources of a significant modulation in CoGeNT rates. The trigger rate is very low (few per hour, including noise triggers), precluding trigger saturation effects. Interference from human activity also seems to be absent (figure \ref{fig:diurnal} and discussion in \cite{neal}). However, an arbitrarily long list of other possibilities can be examined. For instance, the performance of the linear gate present in the triggering channel (figure \ref{fig:electronics}) can be considered. Fluctuations in detector leakage current could in principle alter the preamplifier reset period to the point of creating sufficiently large changes in the 0.2\% trigger dead time induced by the inhibit logic signal (Sec.II-B). For these to mimic a modulation in rate of the $\sim$16\% amplitude reported in \cite{Aal11b}, the detector leakage current and reset period would have to inadvertently vary by a factor of $\sim80$. This would induce changes to the FWHM white parallel electronic noise, dominant for the channel monitored in figure \ref{fig:stability}, by a factor $\sim\sqrt{80}$ \cite{pullia}. These are clearly excluded. In addition to this, linear gate blocking circuitry fluctuations having any other origin would affect all pulses independently of their energy or rise time, an effect not observed \cite{Aal11b}. 

An additional example of an instrumental effect able, in principle, to generate event rate fluctuations is the pulse rise time dependence on crystal temperature described in \cite{bela} for n-type germanium detectors. For the CoGeNT detector, these changes would translate into anti-correlated modulations in surface and bulk event rates, which are not observed, and only for very large seasonal swings in detector temperature of $>\!10$ degrees Celsius. These temperature swings are not expected, given the precautionary 48~hour automatic refills of the Dewar, and the constant LN2 consumption through the year. Ambient temperature at the location of the CoGeNT detector (20.5~$^{\circ}$C) is monitored to be constant within $\pm1$~$^\circ$C, the expected maximum yearly temperature variation in detector and DAQ. In addition to this, the effect is expected to be less noticeable for p-type diodes, which feature considerably better charge mobility than n-type detectors. However, it is worth emphasizing the existence of such subtle instrumental effects, in order to fully appreciate the difficulties involved in obtaining convincing evidence for a dark matter annual modulation signature from any single experiment. A pragmatic approach to this issue is to redesign as much of the DAQ and electronics as possible in all future searches, as planned for the C-4 experiment \cite{inprep}.

\section{Data Selection Cuts}

\begin{figure}[!htbp]
\includegraphics[height=.8\textheight]{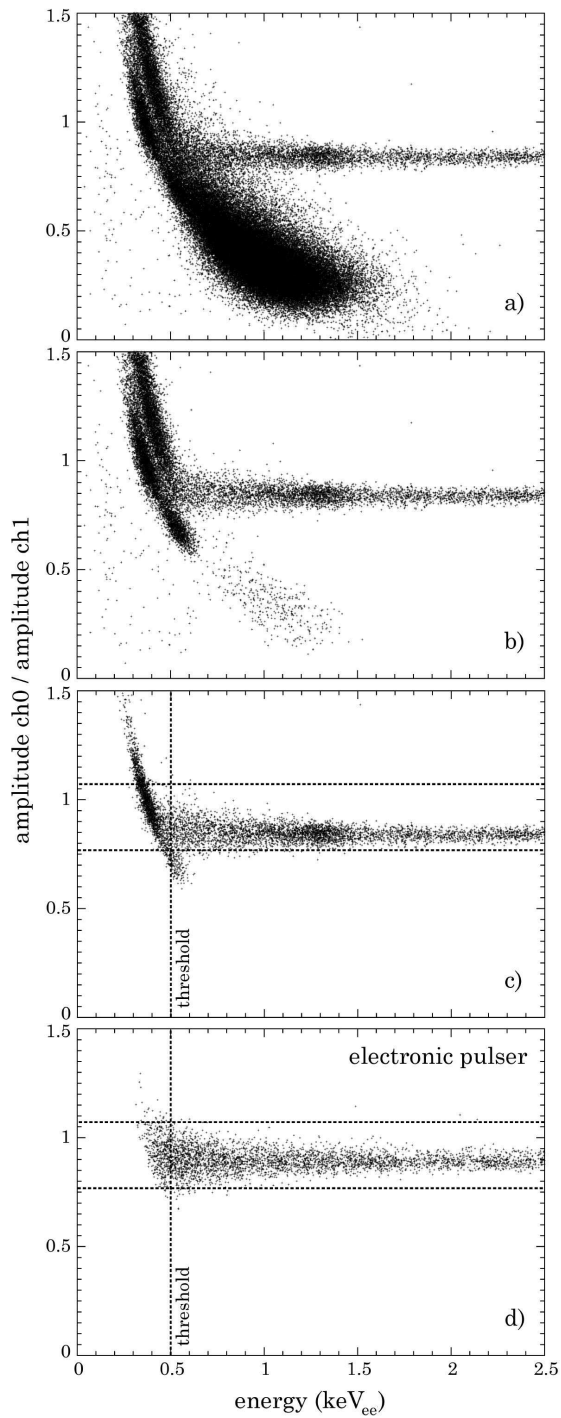}
\caption{\label{fig:analysis} Steps in data selection through the UC analysis pipeline: a) All data including microphonics-intensive periods of LN2 Dewar filling. b) Following removal of LN2 transfer periods and ensuing 10 minutes (boiling in the Dewar lasts a few minutes). No correlated excess of events is observed to extend beyond this 10 min cut. c) Following application of cuts intended to remove anomalous electronic pulses (see text). The boundaries for a final cut using the CH0/CH1 amplitude method in \protect\cite{julio} are shown as horizontal lines. These boundaries are selected to minimize the effect of this cut for both radiation-induced and pulser events, with the exception of a distinct family of residual microphonic events visible as a diagonal band in this panel. d) Fast electronic pulser events prior to any cuts (only the CH0/CH1 amplitude criterion is seen to minimally affect these).}
\end{figure}

\begin{figure}[!htbp]
\includegraphics[height=.22\textheight]{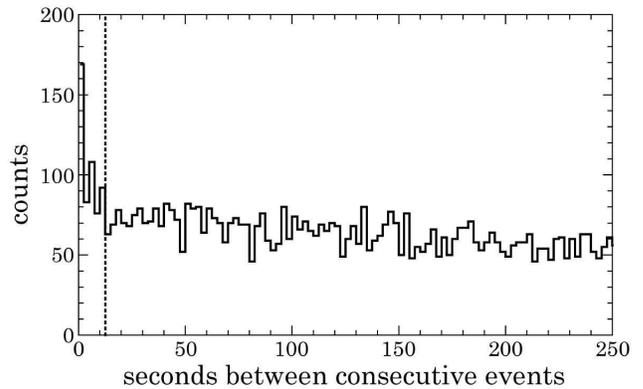}
\caption{\label{fig:bursts} Distribution of time span between consecutive events passing microphonic cuts (see text). A small deviation from a Poisson distribution is observed at t$<$12 s. A large fraction of events in the first bin correspond to the decay of cosmogenic $^{73}$As, involving a short-lived (t$_{1/2}$=0.5 s) excited state \protect\cite{Aal11,phil}.}
\end{figure}

\begin{figure}[!htbp]
\includegraphics[height=.2\textheight]{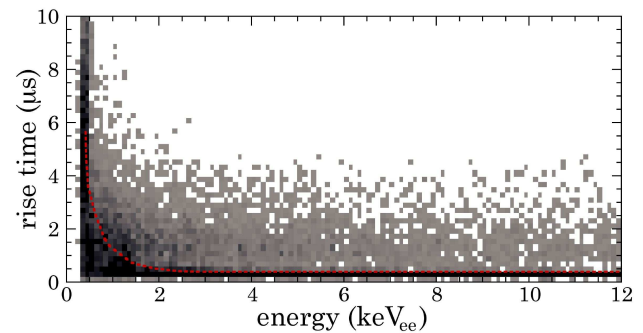}
\caption{\label{fig:grayscale} Grayscale plot showing the distribution of rise time vs.  energy for events passing all other cuts, collected over a 27 month live period for the detector at SUL.  Fast bulk events appear highly concentrated around a $\sim$325 ns rise time, their distribution becoming progressively slower towards zero energy by the effect of electronic noise in preamplifier traces (Sec.IV-B), already visibly affecting the cosmogenic peaks around 1.3 keV. The dotted red line corresponds to the 90\% acceptance boundary for fast electronic pulse events, used for rise time cuts in \protect\cite{Aal11, Aal11b}.}
\end{figure}

\begin{figure}[!htbp]
\includegraphics[height=0.3\textheight]{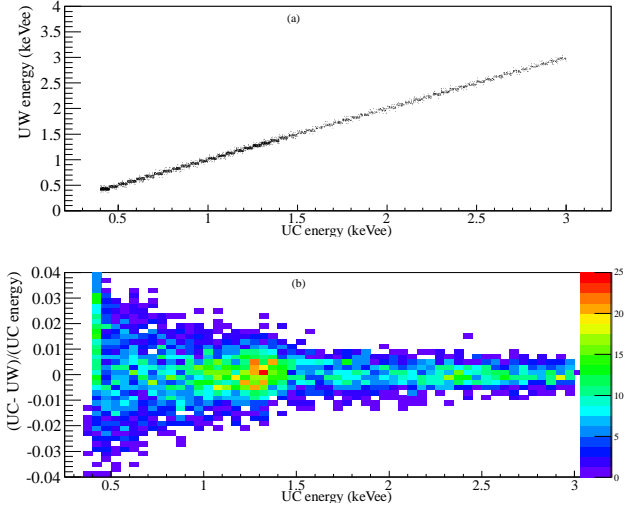}
\caption{\label{fig:EnergyEstimatorCompare} Event-by-event comparison of energy estimators from UC and UW data analysis pipelines (442 day dataset, \cite{Aal11b}).  The top panel shows that the two energy estimators are very well correlated. The bottom panel indicates that the maximum difference between energy estimators is $<$ 4\% above the analysis threshold of 0.5 keV.}
\end{figure}

\begin{figure}[!htbp]
\includegraphics[height=0.3\textheight]{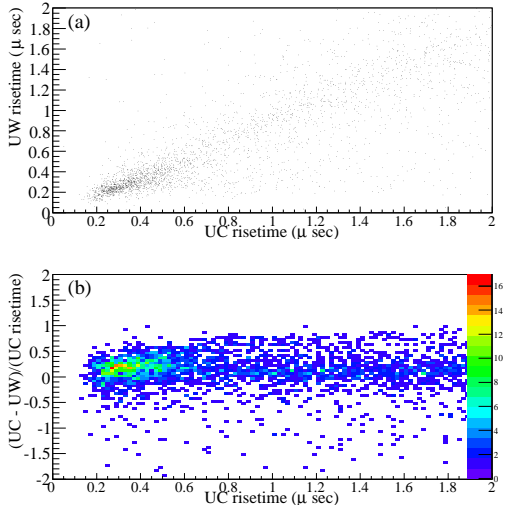}
\caption{\label{fig:RisetimeEstimatorCompare} Event-by-event comparison of rise time estimators from UC and UW data analysis pipelines (442~day dataset, \cite{Aal11b}) in the region 0.5-3.0~keVee.  The top panel (a) shows the correlation between the two estimators. The fractional difference between the two estimators is shown in (b). The two rise time estimators are fairly well correlated, with a disagreement in the classification as fast (bulk) or slow (surface) for only 11\% of the events in the 0.5-3.0~keVee analysis region.  In the region of 0.5-1.0~keVee this disagreement affects 16\% of the events.}
\end{figure}

\begin{figure}[!htbp]
\includegraphics[height=0.3\textheight]{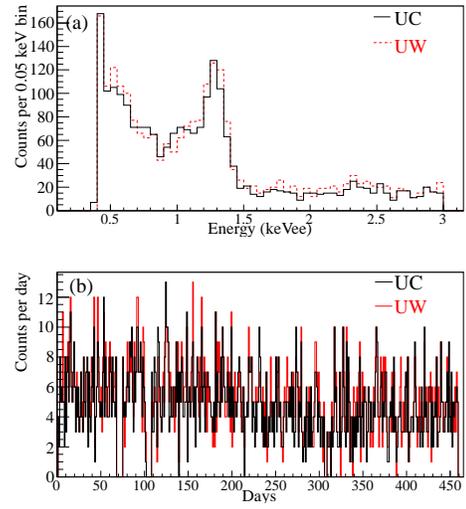}
\caption{\label{fig:RisetimeEstimatorCompare2} Comparison of energy spectra and overall event rate from the UC and UW analysis pipelines.  Panel (a) shows the similar energy spectra obtained following independent data selection and rise time cuts. Panel (b) displays the daily rates in the region 0.5-3.0~keVee for events passing all cuts.}
\end{figure}

\begin{figure}[!htbp]
\includegraphics[height=0.3\textheight]{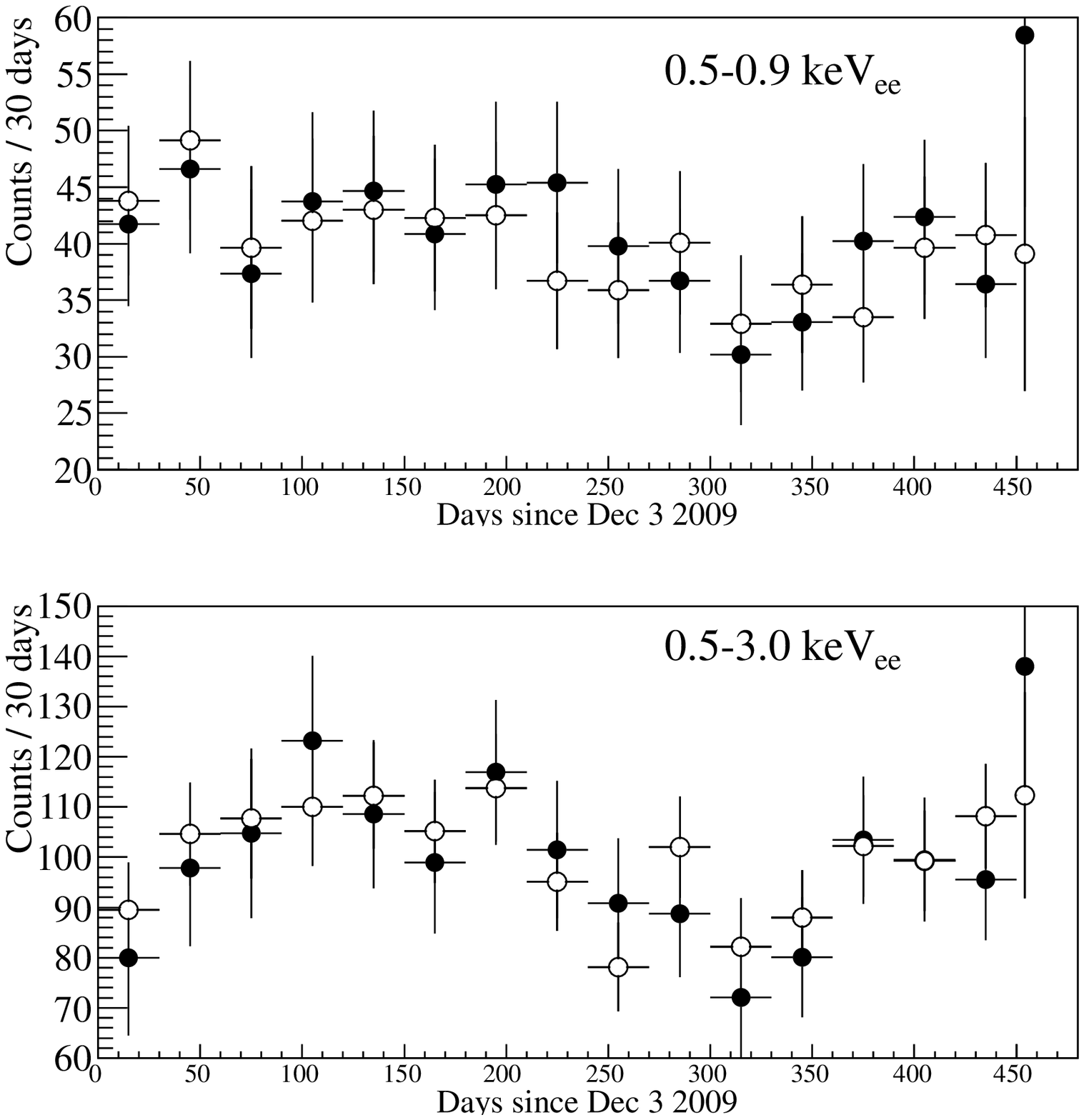}
\caption{\label{fig:modulation} Comparison between irreducible monthly rates in two different energy regions, for the UC (black) and UW (circles) analysis pipelines. The correction for low-energy cosmogenics present in these regions \protect\cite{Aal11b} is applied, and calculated independently for each pipeline.}
\end{figure}

The data acquisition system described in Sec.II-B is designed to exploit a technique detailed in \cite{julio}, able to provide efficient discrimination against low-energy microphonic pulses arising from acoustic or mechanical disturbances to the detector.  In this method, any anomalous preamplifier trace characteristic of a microphonic event is assigned markedly different amplitudes when processed through amplifiers set to dissimilar shaping times (CH0 and CH1 here, figure~\ref{fig:analysis}). An alternative approach to microphonic rejection based on wavelet analysis \cite{igor} was tested. It was found to offer no advantage over that in \cite{julio} for these data, while imposing a considerable penalty on the analysis CPU time. In addition to this microphonic cut, preamplifier traces are screened against deviations from the pattern of a normal radiation-induced pulse (rise time of less than a few $\mu$s, decay time $\sim$50~$\mu$s): several custom data cuts discriminate against sporadic characteristic electronic noise signals (ringing, spikes, reverse polarity pulses from HV micro-discharges, ``telegraph'' noise). These cuts are observed to remove a majority of microphonic pulses on their own, even prior to CH0/CH1 amplitude ratio cuts (figure~\ref{fig:analysis}). As in \cite{julio}, we observe a very small number of microphonic events escaping amplitude ratio cuts. These can be identified by their time correlation, appearing in bunches around times of disturbance. They are removed with an additional time cut (vertical line in figure~\ref{fig:bursts}) that imposes a negligible dead time. 

A final cut selects fast rise time preamplifier pulses, identified with those taking place in the fiducial bulk volume of the crystal, i.e., rejecting the majority of slow, partial charge collection pulses originating in the surface transition layer (Sec.III-C, \cite{Aal11}). This cut is defined by the energy-dependent boundary for 90\%  acceptance of fast electronic pulser signals (figure~\ref{fig:grayscale}, \cite{Aal11}). Pulser scans are used to build an efficiency curve in passing all analysis cuts, used in combination with the trigger efficiency (figure~\ref{fig:triggering}) to generate a modest correction to the energy spectrum \cite{Aal11,Aal11b} (top panel in figure~\ref{fig:steps}).  

Two parallel schemes were developed for CoGeNT data analysis. Both employ independent methods of wavelet denoising on preamplifier traces previous to rise time determination, which also follows separate algorithms. Custom cuts against electronic noise are also independently designed, as well as those for microphonic rejection. Emphasis was placed on avoiding mutual influence between the teams developing these analysis pipelines. The first one, developed at University of Chicago (``UC'') was employed in \cite{Aal08, Aal11, Aal11b}. The second, developed at University of Washington \cite{mike} (``UW'') was used in cross-checking the results in \cite{Aal11, Aal11b}.  There is good event overlap between the two analysis pipelines, with roughly 90\% of the events passing one set of cuts also passing the other. Figures~\ref{fig:EnergyEstimatorCompare},  \ref{fig:RisetimeEstimatorCompare}, \ref{fig:RisetimeEstimatorCompare2}, and  \ref{fig:modulation} display several of the cross-checks performed prior to publication of a search for an annual modulation \cite{Aal11b}. Slighty more events pass the UW risetime cut than the UC risetime cut, by 6.7\%.  Both pipelines generate remarkably close irreducible energy spectra and temporal evolution (figures~\ref{fig:RisetimeEstimatorCompare2} and ~\ref{fig:modulation}). In particular, the possible modulation investigated in \cite{Aal11b} is visible in both lines of analysis (figure~\ref{fig:modulation}). The parameters used for data selection cuts for both pipelines are constant in time, and in the case of the UC pipeline, they were frozen prior to the publication of \cite{Aal11}, implementing a {\it de facto} blind analysis for the larger dataset in \cite{Aal11b}.

\subsection{Cosmic ray veto cuts}

While the CoGeNT detector at SUL incorporates an active muon veto system, no veto cuts are applied to the data in \cite{Aal11,Aal11b}. This is done to avoid introducing any artificial modulation to the event rates arising from fluctuations in the efficiency of this veto or its electronics (recall its setting to single photo-electron detection, which makes it particularly sensitive to such effects). As discussed in this section, it is however possible to make use of this veto to demonstrate that only a negligible fraction of the low-energy events arise from muon-induced radiations, rendering this cut superfluous. This negligible contribution is confirmed by the ($\mu$,n) and ($\mu$,$\gamma$) simulations discussed in Sec.V-A.

\begin{figure}[!htbp]
\includegraphics[height=0.31\textheight]{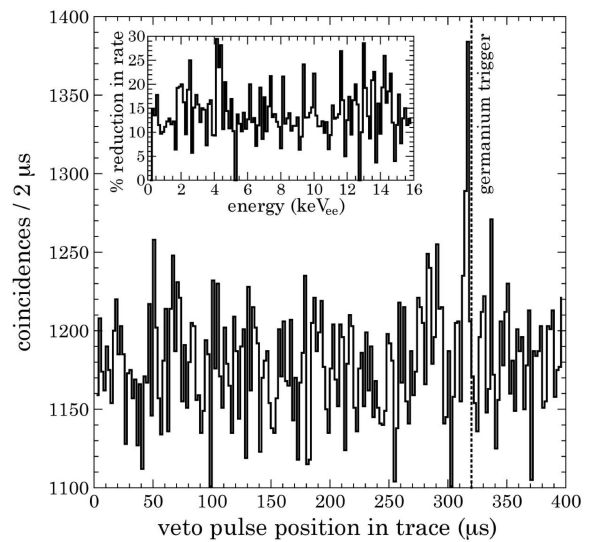}
\caption{\label{fig:muon1} True coincidences between muon veto and PPC appear as an excess above spurious coincidences, displaying the typical delay by a few $\mu$s characteristic of fast neutron straggling. See text for a discussion on the comparison of their rate with that predicted by simulations. Inset: fraction of events removed by a muon veto cut (see text).}
\end{figure}

Operation at single photo-electron sensitivity is required to ensure good efficiency for muon detection from thin (1 cm) scintillator panels, for which a discriminator setting able to separate muon passage from environmental gamma interactions with the veto is not possible. This good efficiency is confirmed by the agreement between the rate of true veto-germanium coincidences (figure~\ref{fig:muon1}) and that predicted by the simulations (Sec.V-A). Specifically, 0.67$\pm$0.12 true coincidences per day were observed during the 442 d of data analyzed in \cite{Aal11b}, whereas 0.77$\pm$0.15 coincidences per day are expected from ($\mu$,n) and ($\mu,\gamma$) simulations. The price to pay for this good muon-detection efficiency is a high veto triggering rate ($\sim$5,000 Hz), which would result in a $\sim$14\% dead time from dominant spurious coincidences were a veto cut applied to the data. It is however evident that the application of the veto coincidence cut would effectively remove a majority of muon-induced events in the germanium detector. 

\begin{figure}[!htbp]
\includegraphics[width=0.34\textheight]{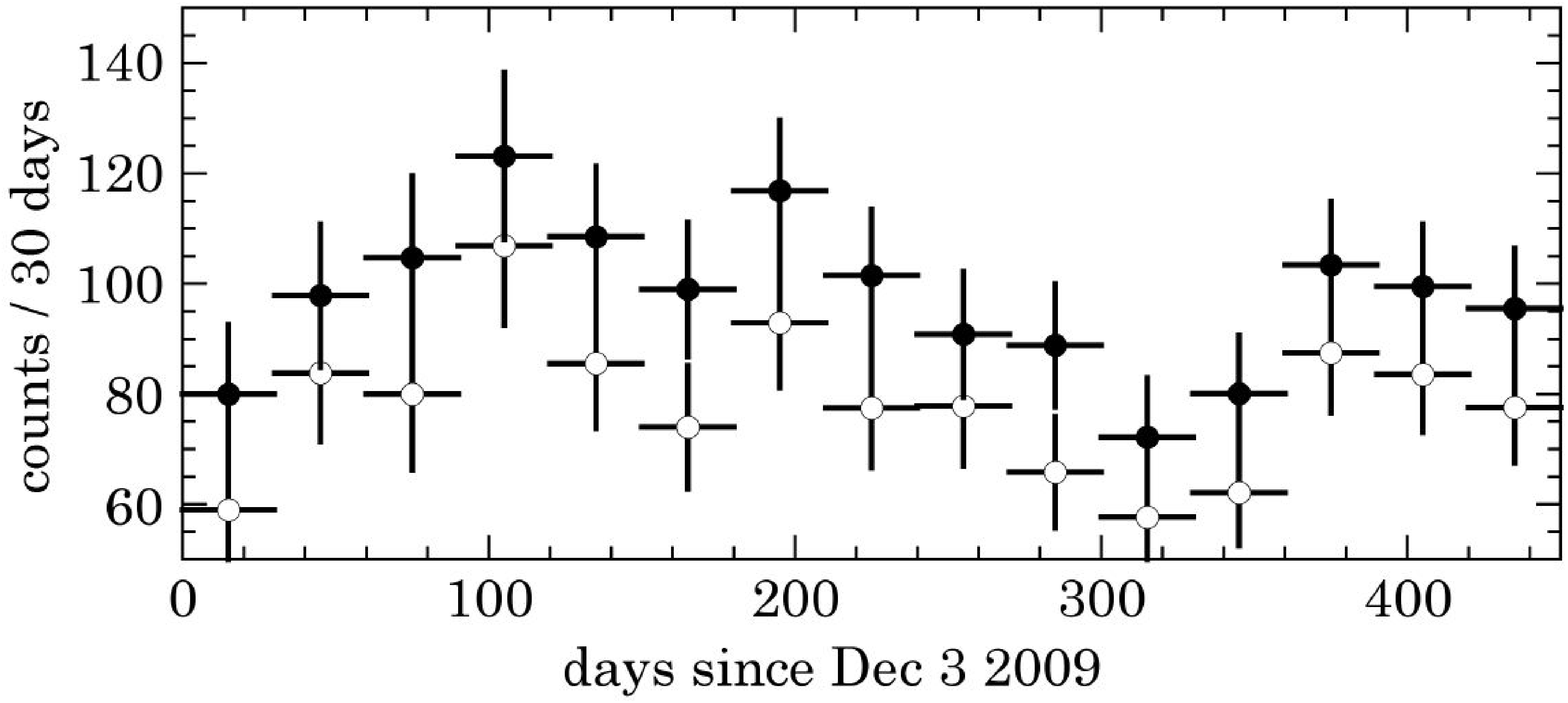}
\caption{\label{fig:muon2} Effect of the application of a veto coincidence cut on the monthly irreducible event rate (see text). White circles incorporate this cut following all other data cuts, as opposed to the inset of figure~\ref{fig:muon1}, where it is applied directly on uncut data. This leads to minor differences in the obtained reduction in event rate. The energy range for this figure is 0.5-3.0~keVee.}
\end{figure}

The inset in figure~\ref{fig:muon1} displays as a function of energy the fraction of events that are removed by application of this cut with a conservative 20 $\mu$s coincidence window. No deviation from the $\sim$14\% rate reduction expected from spurious coincidences is noticeable at low energy, indicating that at maximum a few percent of the spectral rise at low energy observed in \cite{Aal11,Aal11b} can be due to muon-induced events. A similar conclusion is derived from the simulations in Sec.V-A. As expected, the application of the veto cut simply decreases the irreducible event rate by this $\sim$14\% fraction, not altering the possible modulation investigated in \cite{Aal11b} (figure~\ref{fig:muon2}). In Sec.V-A we will conclude that the muon-induced modulation amplitude expected for CoGeNT at SUL is of O(0.1)\%. Separately, the MINOS collaboration finds a three-sigma inconsistency between the phases of their measured modulation in muon flux at SUL, and that observed in CoGeNT data \cite{minosmod}. 

\subsection{Uncertainties in the rejection of surface events}

As discussed in \cite{Aal11,Aal11b} and visible in figure~\ref{fig:grayscale}, the ability to discriminate between fast rise time (bulk) and slow rise time (surface) events is progressively diminished for energies approaching the 0.5~keVee threshold. When the amplitude of a preamplifier pulse becomes close to the amplitude of the circuit's electronic noise variations, an accurate measurement of rise time becomes more difficult to perform, even after wavelet denoising. Determining the bulk-event signal acceptance (SA) is straightforward when electronic pulser signals are identified to be a close replica of fast radiation-induced events in the bulk of the crystal \cite{Aal11}. In the analysis described in this section this SA is kept at an energy-independent 90\% (red dotted line in figure~\ref{fig:grayscale}),  as in \cite{Aal11,Aal11b}. Using an additional 12 months of exposure beyond the dataset in \cite{Aal11b},  we can finally attempt the exercise of calculating the surface event background rejection (BR) as a function of energy. It must be emphasized that the resulting correction (the true fraction of bulk events in those passing all cuts, figure~\ref{fig:surface_correction}) can only be applied to the irreducible energy spectrum, and not to individual pulses on an event-by-event basis, similar to the case of low-energy nuclear and electron recoil discrimination in sodium iodide detectors \cite{smith}.

\begin{figure}[!htbp]
\includegraphics[width=0.36\textheight]{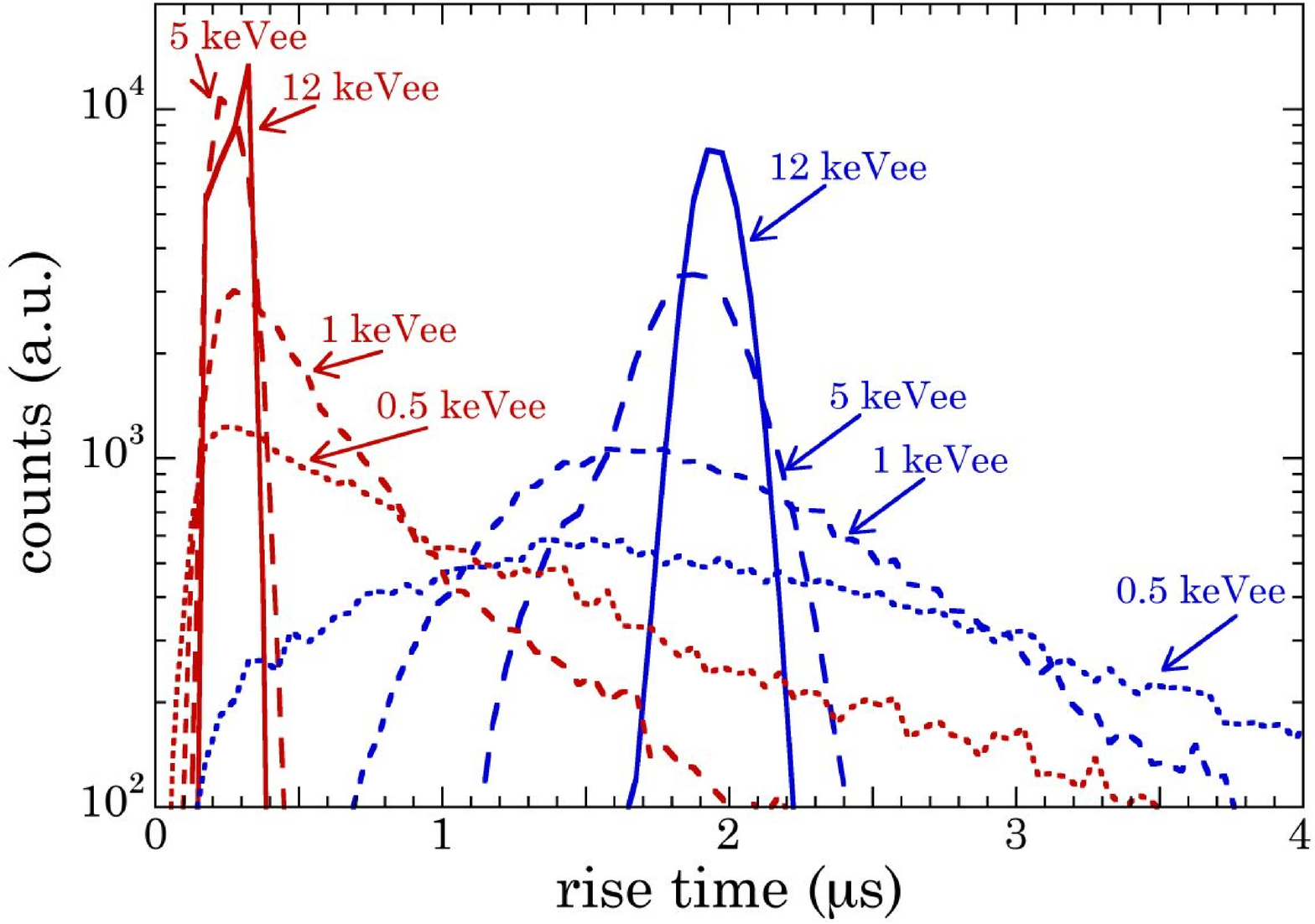}
\caption{\label{fig:surface_simulations} Simulated preamplifier pulses with an initial rise time of 325 ns, representing ideal fast (bulk) events, are convoluted with electronic noise and treated with the same wavelet denoising and rise time measurement algorithms applied to real events. This electronic noise is grafted directly from pre-trigger preamplifier traces taken from real detector events, leading to perfect modeling of the noise frequency spectrum. The resulting rise time distributions are represented by red curves, labelled by their energy equivalent. The same is repeated for typical slow (surface) pulses with a rise time of 2~$\mu$s, generating the blue curves. Each simulation contains 35k events. These simulations provide a qualitative understanding of the behavior observed in figure~\ref{fig:grayscale}.}
\end{figure}

\begin{figure}[!htbp]
\includegraphics[width=0.35\textheight]{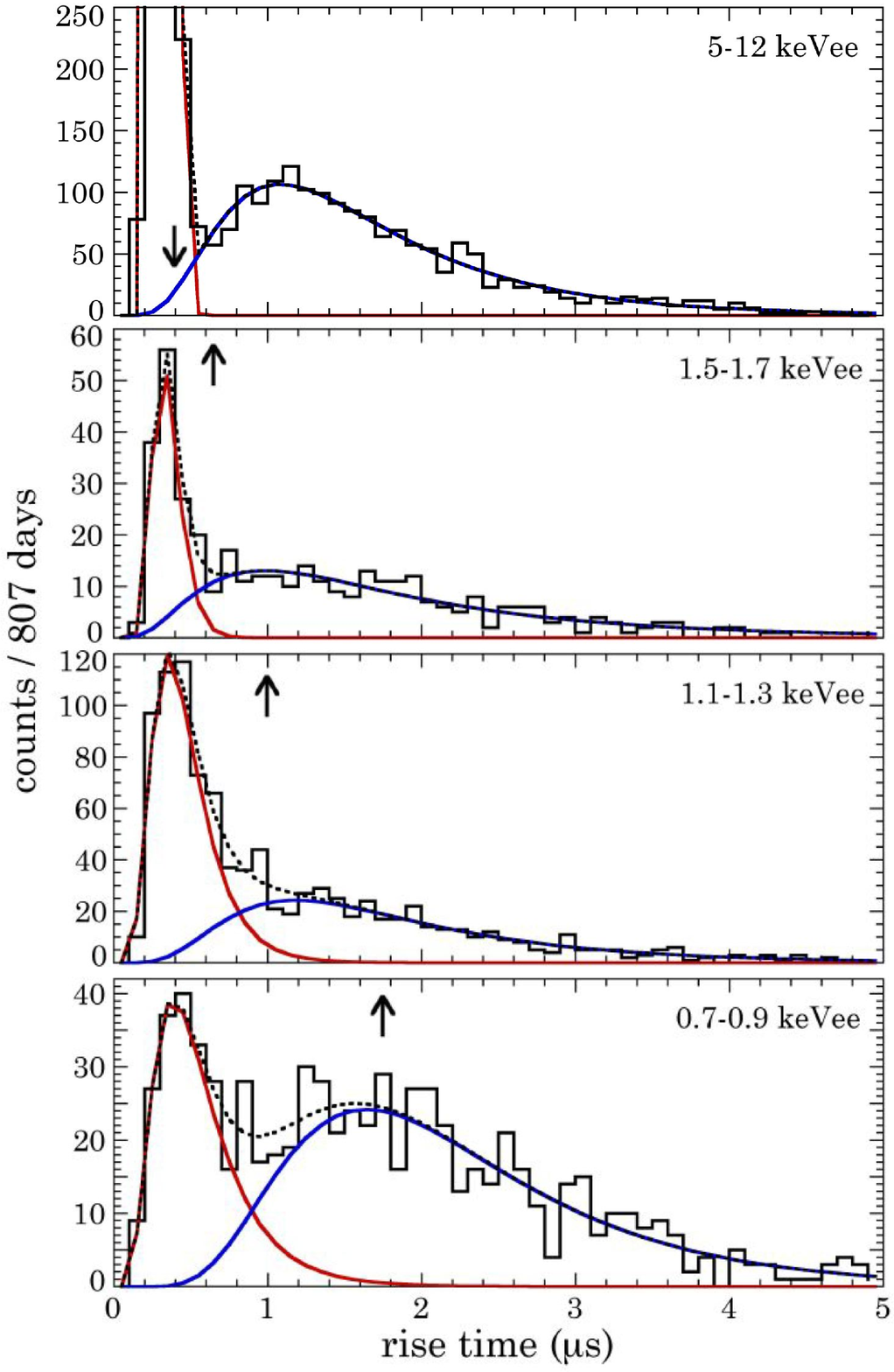}
\caption{\label{fig:surface_fits} Example rise time distributions for events falling within discrete energy bins, from a 27~month exposure of the CoGeNT detector at SUL. These are fitted by two log-normal distributions with free parameters, corresponding to slow rise time surface events (blue) and fast rise time bulk events (red). Small vertical arrows point at the location of the 90\% C.L. fast signal acceptance boundary dictated by electronic pulser calibrations (dotted red line in figure~\ref{fig:grayscale}). A contamination of the events passing this cut by unrejected surface events progresses as energy decreases (see text).  }
\end{figure}

\begin{figure}[!htbp]
\includegraphics[width=0.35\textheight]{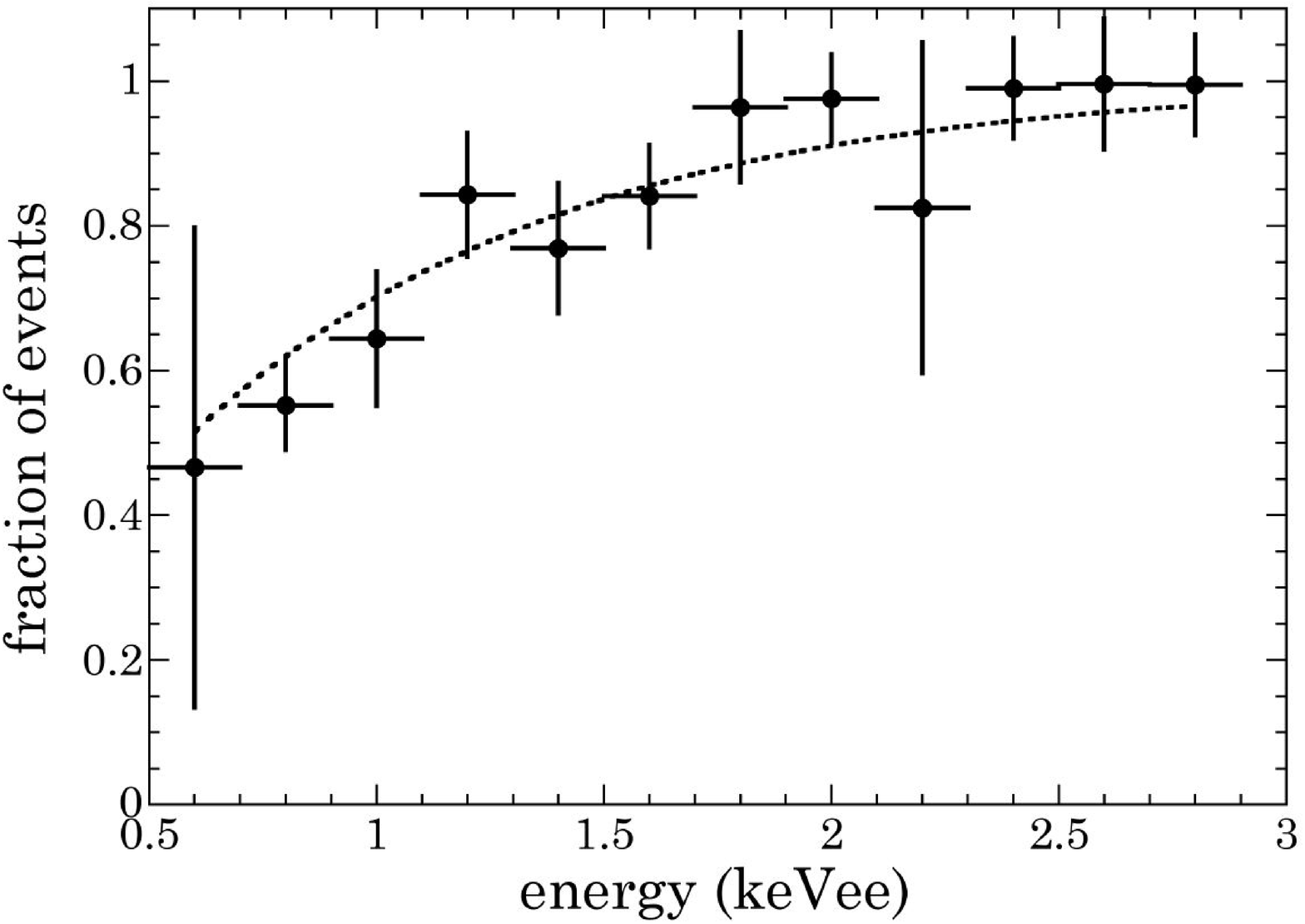}
\caption{\label{fig:surface_correction} Fraction of events passing the 90\% fast signal acceptance cut (pulser cut, dotted red line in figure~\ref{fig:grayscale}) identified as true bulk events via the analysis discussed in Sec.IV-B. Alternatively defined, its complement is the fraction of events passing the pulser cut that are in actuality misidentified surface events (see figure~\ref{fig:surface_fits}). The dotted line is a fit with functional form $1-e^{-a\cdot E(keVee)}$, with $a=1.21\pm0.11$. Error bars are extracted from the uncertainties in fits like those exemplified in figure~\ref{fig:surface_fits}.}
\end{figure}

\begin{figure}[!htbp]
\includegraphics[width=0.35\textheight]{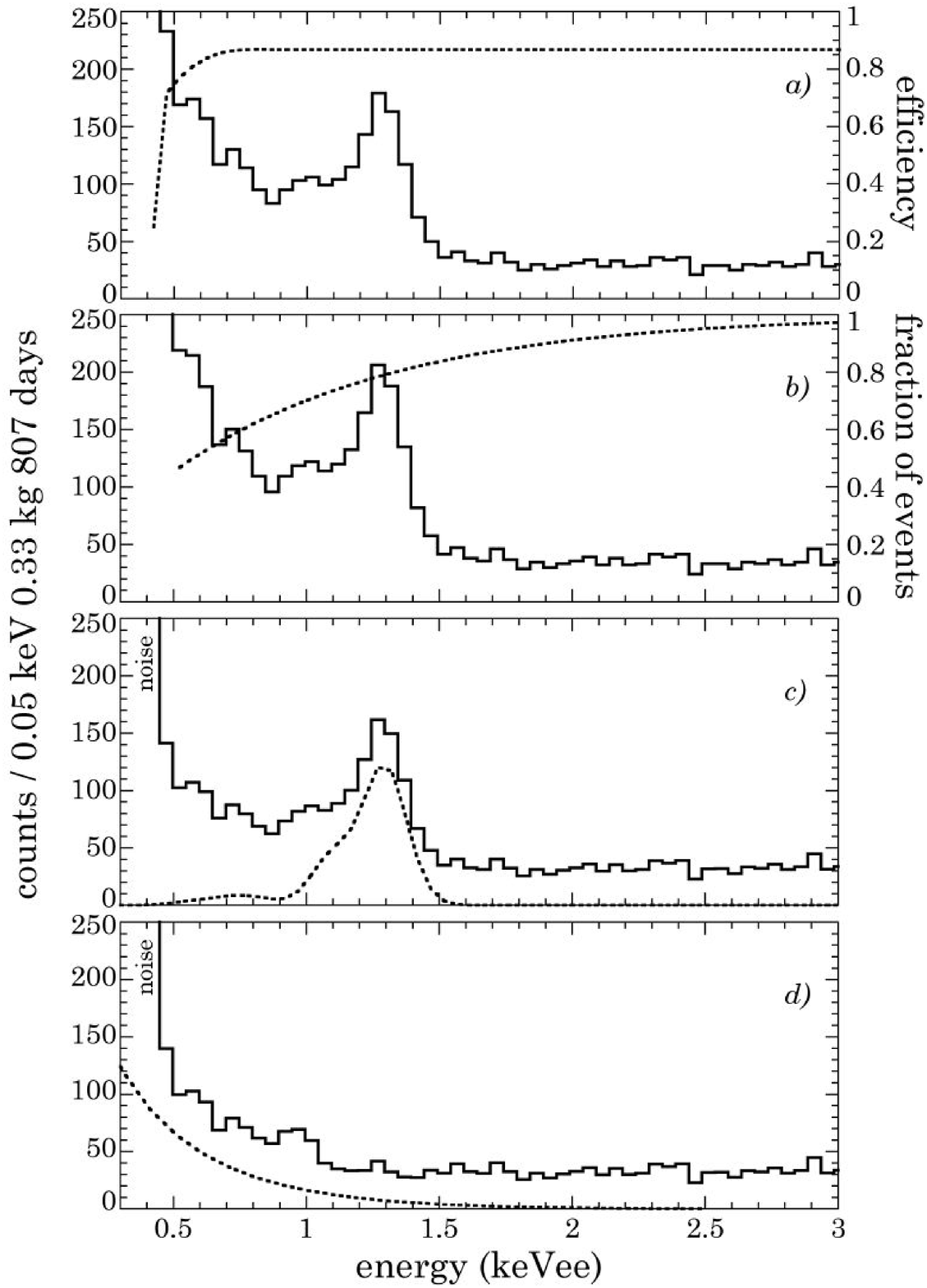}
\caption{\label{fig:steps} Steps in the treatment of a low-energy CoGeNT spectrum. a) Spectrum following data selection cuts (Sec.IV), including 90\% fast signal acceptance cuts from pulser calibrations (dotted red line in figure~\ref{fig:grayscale}) \cite{Aal11,Aal11b}. This spectrum is nominally composed by a majority of bulk events. Overimposed is the combined trigger and background cut efficiency. This efficiency is derived from high-statistics pulser runs (figures~\ref{fig:triggering} and \ref{fig:analysis}), resulting in a negligible associated uncertainty. b) Spectrum following this trigger plus cut efficiency correction. Overimposed is the residual surface event correction. This correction and its associated uncertainty can be found in figure~\ref{fig:surface_correction}. c) Spectrum following this surface event contamination correction. Overimposed is the predicted cosmogenic background contribution, reduced by 10\% as in \cite{Aal11b}. The modest uncertainties associated to this prediction, dominated by present knowledge of L/K shell electron capture ratios, are discussed in \cite{Aal11b}. d) Irreducible spectrum of bulk events, now devoid of surface and cosmogenic contaminations \cite{consistency,gerbier1}. Overimposed is the expected signal from a m$_{\chi} = 8.2$~GeV/c$^{2}$, $\sigma_{SI} = 2.2 \times 10^{-41}$~cm$^{2}$ WIMP, corresponding to the best-fit to a possible nuclear recoil excess in CDMS germanium detector data \cite{ourcdms}. A bump-like feature around 0.95~keVee is absent in the alternative UW analysis shown in figure~\ref{fig:RisetimeEstimatorCompare2} and is therefore likely merely a fluctuation.}
\end{figure}

\begin{figure}[!htbp]
\includegraphics[width=0.35\textheight]{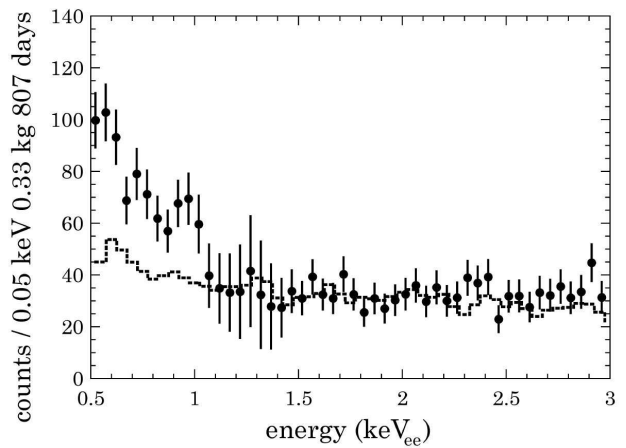}
\caption{\label{fig:overlay} Irreducible spectrum of bulk events (points) showing cumulative uncertainties from the corrective steps discussed in figure~\ref{fig:steps}. The simulated total background spectrum from Sec.V is shown as a histogram, scaled to the larger exposure in this figure, and corrected for the combined trigger and background cut efficiency.}
\end{figure}

\begin{figure}[!htbp]
\includegraphics[width=0.35\textheight]{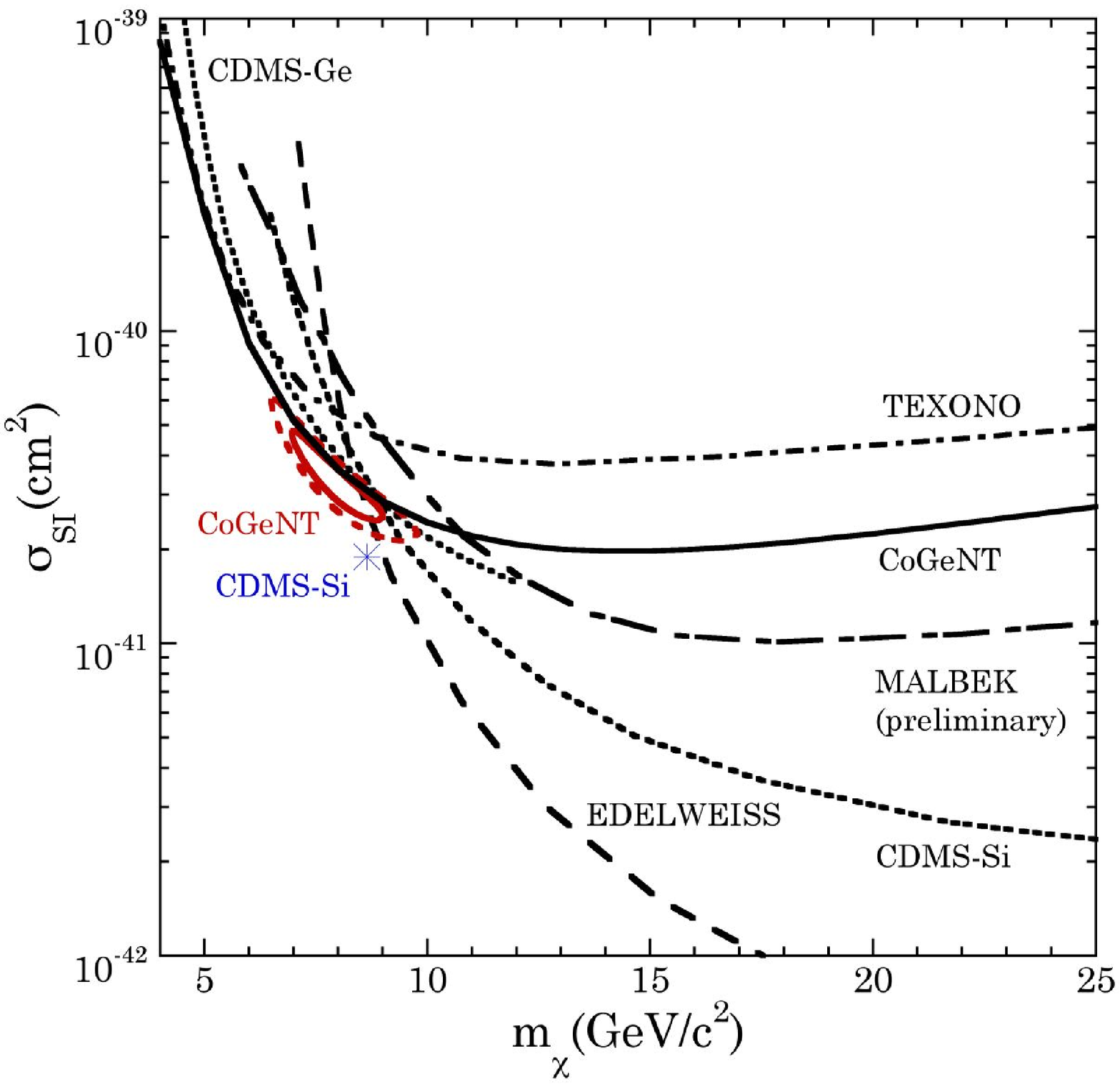}
\caption{\label{fig:roi} 90\% C.L. WIMP limits extracted from the irreducible bulk event spectrum in figure~\ref{fig:overlay}, placed in the context of other low-threshold detectors. A Maxwellian galactic halo is assumed, with local parameters
$v_{0}=$220~km/s,  $v_{esc}=$550 km/s, $\rho=$0.3 GeV/c$^{2}$cm$^{3}$. A ROI (red solid 90\% C.L., red dashed 99\% C.L.) can be extracted if a WIMP origin is assigned to the rise in the spectrum. This ROI includes the cumulative uncertainties shown in figure~\ref{fig:overlay}, and allows for a flat background component,  independent of energy, in addition to a WIMP signal. The reader is referred to \protect\cite{kelso} for a discussion on astrophysical uncertainties not included here (see also \protect\cite{Aal11b,gerbier1}). This ROI partially overlaps with another one, not shown here for clarity, extracted from a possible excess of low-energy nuclear recoils in CDMS germanium data \protect\cite{ourcdms}. A best-fit to that possible excess is shown in the bottom panel of figure~\ref{fig:steps}.  Recent low-mass WIMP limits from CDMS-Ge \protect\cite{prevcdms}, EDELWEISS \protect\cite{edelweiss}, TEXONO \protect\cite{texono}, MALBEK \protect\cite{malbek}, and CDMS-Si \protect\cite{cdmssi} are indicated. A blue asterisk indicates the centroid within a large ROI generated by an excess of three nuclear-recoils in CDMS silicon detector data  \protect\cite{cdmssi}.}
\end{figure}

In the ideal situation where all radiation sources affecting the detector were known in intensity, radioisotope and location, including surface activities, it might be possible to consider a simulation able to predict the exact distribution of pulse rise times as a function of measured energy. This simulation would also require a precise knowledge of the surface layer structure estimated in Sec.III-C (charge collection efficiency and pulse rise time should correlate within the transition region \cite{sakai}), and modeling of the ensuing processes of charge transport and electronic signal generation. This approach is particularly unrealistic when dealing with few keVee energy depositions.  Calibrations using external gamma sources are of value in understanding the structure and effect of the transition layer \cite{Aal11},  but cannot replicate the exact distribution of events in rise time vs. energy during physics runs,  which is specific of the particular environmental radiation field affecting a PPC.  

An alternative route ensues from a study of simulated preamplifier pulses, as described in figure~\ref{fig:surface_simulations}. These provide a qualitative understanding of the blending in rise time of surface and bulk events as energy decreases. It is also observed that all simulated rise time distributions can be described by log-normal probability distributions. A next step is to divide the large (27 month) dataset accumulated up to June 2012 into discrete energy bins for events passing all cuts, but prior to any discrimination based on rise time (figure~\ref{fig:surface_fits}). This large exposure allows study of the evolution of these two families of events as a function of energy. Surface and bulk events are observed to form two distinct distributions for energies above a few keVee (top panel in figure~\ref{fig:surface_fits}), where the impact of the electronic noise on rise time measurements is minimal (figure~\ref{fig:surface_simulations}). A progressive mixing of the two distributions, expected qualitatively from the simulations, is observed to take place at lower energies (figure~\ref{fig:surface_fits}). This results in a contamination with unrejected surface (slow) events of the energy spectrum of pulses passing the 90\% C.L. fast signal acceptance cut derived from electronic pulser calibrations (figure~\ref{fig:grayscale}). The magnitude of this contamination (figure~\ref{fig:surface_correction}) can be derived from the fits to the rise time distributions shown in figure~\ref{fig:surface_fits}, and to others like them. The electronic pulser cut (vertical arrows in figure~\ref{fig:surface_fits}) correctly approximates the $\sim$90\% boundary to the fitted fast pulse distributions (shown in red), confirming that bulk event SA can be correctly estimated using the electronic pulser method.

These fits reveal two significant trends, both visible in figure~\ref{fig:surface_fits}: first, the mean of the slow pulse distribution is seen to drift towards slower rise times with decreasing energy, an effect already observed in surface irradiations of PPCs using  $^{241}$Am gammas \cite{Aal11,ryan}. Second, the standard deviation of the fitted fast pulse distribution (i.e., its broadening towards slower rise times) is noticed to increase with decreasing energy, in good qualitative agreement with the behavior expected from simulated pulses (figure~\ref{fig:surface_simulations}).

Figure~\ref{fig:steps} summarizes the steps necessary in the treatment of CoGeNT low-energy data, leading to an irreducible spectrum of events taking place within the bulk of the crystal, devoid of surface events and cosmogenic backgrounds \cite{consistency}. As discussed in the following section, the exponential excess observed at low energy is hard to understand based on presently known radioactive backgrounds.  Figure~\ref{fig:overlay} shows the irreducible spectrum of bulk events including the uncertainties discussed in figure \ref{fig:steps}, overlayed with the total background estimate from Sec.V, pointing at an excess of events above the background estimate.  Figure \ref{fig:roi} displays WIMP exclusion limits that can be extracted from this irreducible spectrum, compared to those from other low-threshold detectors. The figure includes a region of interest (ROI) generated when assuming a WIMP origin for the low-energy exponential excess.  

Best-fit distributions like those in figure~\ref{fig:surface_fits} point at the possibility of obtaining $\sim$45\% BR of surface events for a 90\% SA of bulk events at 0.5~keVee threshold, rapidly rising to $\sim$90\% BR at 1.0 keV, for the same 90\% SA. A pragmatic approach to improving this event-by-event separation between surface and bulk events, is to tackle the origin of the issue, i.e., to further improve the electronic noise of PPCs. A path towards achieving this within the C-4 experiment is delineated in \cite{inprep}. In the mean time, the large exposure collected by the PPC at SUL should allow a refined weighted likelihood annual modulation analysis, in which the rise time of individual events provides a probability for their belonging to the surface or bulk categories (figure~\ref{fig:surface_fits}). This analysis is in preparation.

\section{Background Studies}

\begin{figure}[!htbp]
\includegraphics[height=0.24\textheight]{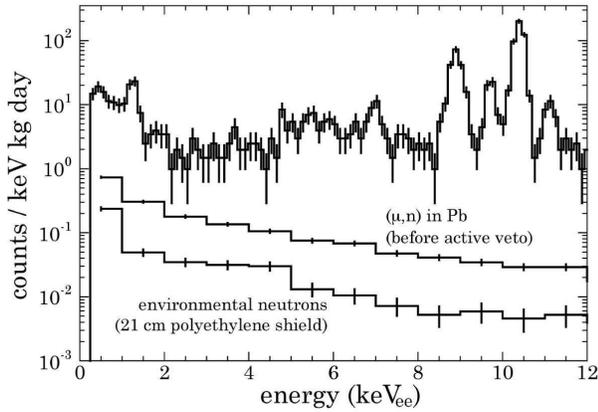}
\caption{\label{fig:mcnp} MCNP-Polimi neutron simulations compared with an early spectrum from the CoGeNT detector at SUL (see text).}
\end{figure}

The present understanding of backgrounds affecting the CoGeNT detector at SUL is described in this section, including contributions from neutrons, both muon-induced and also for those arising from natural radioactivity in the SUL cavern. Early calculations for these made use of MCNP-Polimi \cite{polimi} simulations, NJOY-generated germanium cross-section libraries, muon-induced neutron yields and emission spectra exclusively from the (dominant) lead-shielding target as in \cite{ming,spectra}, and SUL cavern neutron fluxes from \cite{cdmsn}. These are shown in figure~\ref{fig:mcnp}. Fair agreement (better than 50\% overall) was found between these and subsequent GEANT \cite{G} simulations, which however include muon-induced neutron production in the full shield assembly and cavern walls, and are able to track the (subdominant) electromagnetic component from muon interactions. The rest of this chapter describes these more comprehensive GEANT simulations. 

\subsection{Neutrons}

\subsubsection{Muon-Induced Neutrons}

\begin{figure}[!htbp]
\includegraphics[height=0.23\textheight]{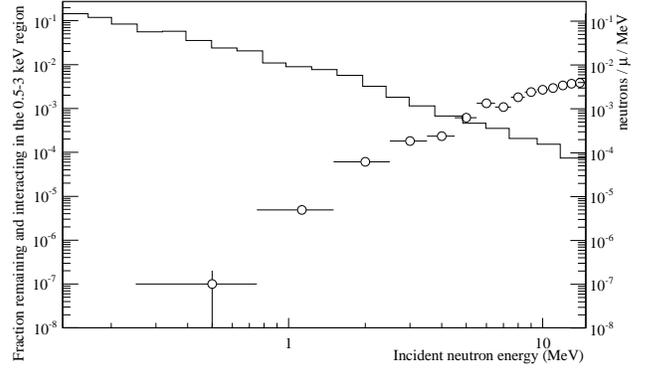}
\caption{\label{fig:neutronfraction} Fraction of external ($\mu$,n) cavern neutrons giving rise to energy depositions in the 0.5-3.0~keVee energy window of the CoGeNT detector at SUL, as a function of incident neutron energy, derived from a Monte Carlo simulation (open circles).  Also shown, using the right-hand scale, is the emission energy spectrum for these neutrons, taken from \cite{Araujo} (histogram).}
\end{figure}

The muon-induced neutron background can be broken up into two components:  those produced by muon interactions in the cavern walls, and those generated by interactions in the CoGeNT shielding materials.  The energy spectrum of external ($\mu$,n) cavern neutrons was taken from \cite{Araujo}.  Figure \ref{fig:neutronfraction} shows the fraction of these neutrons making it through the shielding and depositing energy in the germanium detector, as a function of incident neutron energy.  The same figure shows the input neutron energy distribution taken from \cite{Araujo} in units of neutrons / $\mu$ / MeV.   Convolving the two distributions, taking into account the muon flux at SUL, and integrating over all neutron energies gives an upper limit of 1.4 external muon-induced neutrons depositing energy in the 0.5-3.0~keVee window for the entire 442~day CoGeNT dataset in \cite{Aal11b}.

\begin{figure}[!htbp]
\includegraphics[height=0.27\textheight]{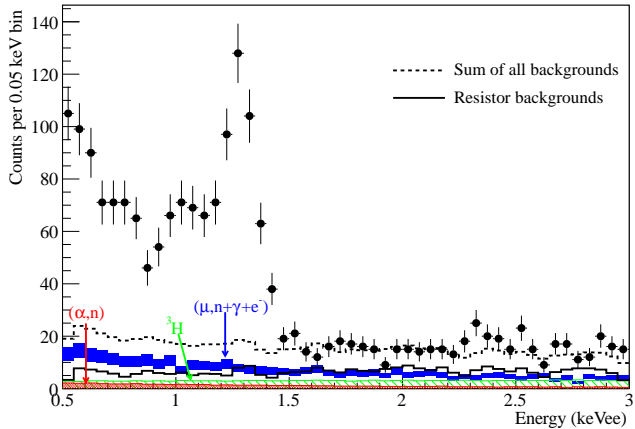}
\caption{\label{fig:muoninduced} Deposited energy spectra from all known backgrounds in the CoGeNT detector at SUL, compared to the 442 d of data in \protect\cite{Aal11b}.  An unidentified low-energy excess and L-shell EC cosmogenic contributions are visible \protect\cite{Aal11, Aal11b}. The corrections in figure~\ref{fig:steps} reduce this excess by $\sim$30\% at 0.5~keVee. The blue band represents the sum of muon-induced backgrounds (Sec.V-A1), the green hatched band is a conservative upper limit to the background from cosmogenic $^{3}$H (Sec.V-B),  and the red band is from ($\alpha$,n) natural radioactivity in cavern walls (Sec.V-A2).  The solid line represents the background distribution from the $^{238}$U and $^{232}$Th chains as well as $^{40}$K contamination in the front-end resistors, estimated in Sec.V-D2.  The dashed line is the sum of all background contributions. Contributions from bremsstrahlung from $^{210}$Pb in the inner lead shield (Sec.II-A) and radioactivity from cryostat parts (Sec.V-D1) are found to contribute negligibly. }
\end{figure}

The largest contribution from neutrons to CoGeNT events arises from spallation neutrons produced by muons traversing the CoGeNT shielding.  Their simulation uses as input the energy and angular distribution given by \cite{ming}.  This simulation also keeps track of electrons, positrons, and gammas produced along the muon track through pair production, subsequent positron annihilation, and bremsstrahlung. Figure \ref{fig:muoninduced} shows the simulated energy deposition of these muon-induced events (blue band) compared to CoGeNT data.  The estimated number of muon-induced events in the 0.5-3.0~keVee region for the 442~day CoGeNT dataset is 339~$\pm$~68.  Only about 8\% of these events involve electron or gamma interactions with the detector, the rest being mediated by neutrons.

Both MCNP-Polimi and GEANT simulations point at less than 10\% of the irreducible rate at threshold in CoGeNT having an origin in ($\mu$,n) sources, an estimate confirmed by the separate muon-veto considerations discussed in Sec.IV-A. The MINOS experiment at the same location provides an accurate measurement of the magnitude of seasonal fluctuations in underground muon flux, limited to less than $\pm$1.5\% \cite{goodm,minos}. Any muon-induced modulation is therefore expected to be of a negligible O(0.1)\% for the present CoGeNT detector. Muons at SUL exhibit a maximum rate on July 9th \cite{minos}, in tension with the best-fit modulation phase found in \cite{Aal11b}. The reader is referred to recent studies \cite{muonstudies} pointing at similar conclusions. Of special relevance is work recently performed by the MINOS collaboration \cite{minosmod}, leading to conclusions similar to those presented here.

\subsubsection{Fission and ($\alpha$,n) neutrons}

The flux of ($\alpha$,n) neutrons from radioactivity in the cavern rock is much higher than that of neutrons produced through muon spallation in the rock.  Cavern  ($\alpha$,n) neutrons were simulated using the energy distribution and flux in \cite{ming}.  The contribution of these cavern ($\alpha$,n) neutrons to the low-energy CoGeNT spectrum is shown in figure \ref{fig:muoninduced} (red band).

The high-density polyethylene (HDPE) in the outer layer of the CoGeNT shielding is known to have relatively high levels of $^{238}$U and $^{232}$Th contamination.  These $^{238}$U and $^{232}$Th concentrations were measured for HDPE samples at SNOLAB, finding 115$\pm$5 mBq/kg and 80$\pm$4 mBq/kg, respectively.  $^{238}$U has a small spontaneous fission (SF) branching ratio with an average multiplicity per fission of 2.07 \cite{Axton}.  Neutrons from this source depositing energy in the 0.5-3.0~keVee region of the spectrum are estimated to be just 17.7$\pm$7.2 for the entire 442~day data set.  An isotope of carbon, $^{13}$C,  has a 1.07\% natural abundance and a non-negligible cross-section for the ($\alpha$,n) reaction at $\alpha$ energies emitted by the U and Th decay chains.  The HDPE is therefore a weak source of ($\alpha$,n) neutrons.  The neutron production from ($\alpha$,n) in HDPE was scaled from a SOURCES \cite{sources} calculation for plastic material \cite{Perry}.  The number of ($\alpha$,n) neutron-induced events in the CoGeNT data set from $^{238}$U and $^{232}$Th in HDPE was determined to be a negligible $<$~0.02 and $<$~0.01, respectively.  Table \ref{tab:neutronsources} summarizes the contributions from the various sources of neutrons in the 442 day CoGeNT data set.  The lead surrounding the detector is also a weak source of fission neutrons.  The $^{238}$U concentration in lead has been measured at SNOLAB to be 0.41$\pm$0.17 mBq/kg.  This results in $<$ 0.5 events from $^{238}$U fission in lead for the entire CoGeNT data set.

\begin{table}[b]
\caption{\label{tab:neutronsources} Summary of backgrounds in a 442 day CoGeNT data set, from various sources investigated.}
\begin{ruledtabular}
\begin{tabular}{cc}
\textrm{Source}&
\textrm{Number of events}\\
\colrule
Cavern muon-induced neutrons & $<$1.4 \\
Cavern ($\alpha$,n) neutrons & $<$54 \\
Muon-induced events in shielding & 339$\pm$68\\
$^{238}$U fission in HDPE & 17.7$\pm$7.2\\
($\alpha$,n) from $^{238}$U in HDPE & $<$0.02 \\
($\alpha$,n) from $^{232}$Th in HDPE & $<$0.01 \\
$^{3}$H  in the Ge detector & $<$150 \\
 $^{238}$U and $^{232}$Th in Cu shield & $\sim$9 \\
$^{238}$U,$^{232}$Th, and $^{40}$K in resistors & $\sim$324 \\
\end{tabular}
\end{ruledtabular}
\end{table}

\subsection{Cosmogenic Backgrounds in Germanium}

Tritium can be produced via neutron spallation of the various natural germanium isotopes.  Most of the $^{3}$H production occurs at the surface of the Earth where the fast neutron flux is much higher than underground.  Tritium has a half-life of 12.3 years, which means its reduction over the lifetime of the experiment is small.  Its beta decay is a potential background for CoGeNT, given its modest end-point energy of 18.6 keV.  Using the $^{3}$H production rate in \cite{Elliott} and \cite{Morales} and assuming an overly conservative two years of sea-level exposure for the crystal, an upper limit of $<$150 $^{3}$H decay events was extracted for the CoGeNT data set. While this number would present a significant background, the energy spectrum of the $^{3}$H events is relatively flat over the 0.5-3.0~keVee analysis region and does not provide for the excess observed at low energies.  Figure ~\ref{fig:muoninduced} shows the upper limit to the contribution from $^{3}$H decays (shaded green) in the analysis region, compared to the data.

All other sufficiently long-lived cosmogenic radioisotopes of germanium produce monochromatic energy depositions at low energy \cite{Aal11,Aal11b,collarthesis}, or have endpoints large enough not to be able to contribute significantly in the few keVee region. The fraction of these taking place in the transition surface layer might however lead to an accumulation of partial charge depositions at energies below the cosmogenic peaks, even if most of these events should in principle be rejected by the rise time cut. That this accumulation is indeed negligible can be ascertained by the lack of correlation between the relatively constant rates shown in figure~\ref{fig:modulation} and the much larger change under the dominant 10.3 keV cosmogenic peak, which reduced its activity from $\sim$500 counts/month to $\sim$150 counts/month over the same period of time. 

\begin{figure}[!htbp]
\includegraphics[height=0.26\textheight]{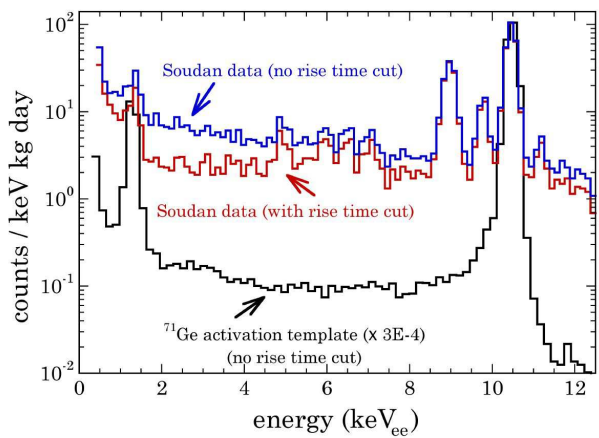}
\caption{\label{fig:ge71} Negligible upper-limit to the contribution from cosmogenic activity in the near-threshold energy region of the CoGeNT detector at SUL (see text).}
\end{figure}

An episode of intense thermal neutron activation of $^{71}$Ge in a PPC with identical characteristics to that operating at SUL, related in \cite{Aal11}, provides additional confirmation that this possible source of background is small. Figure~\ref{fig:ge71} shows the spectrum acquired during the first few days following this thermal neutron activation. The data were taken at the San Onofre nuclear plant at a depth of 30~m.w.e., inside a large passive shield and triple active veto. The initial $^{71}$Ge decay rate under the 10.3~keV peak was very high, at $\sim$0.3~Bq. The low-energy $^{71}$Ge spectral template shown in the figure was therefore entirely dominated by the response to this activation, with the counting rate below 10~keVee dropping by several orders of magnitude over the ensuing weeks, to stabilize at a factor of just a few above the rate observed at SUL. Once the $^{71}$Ge activation template is normalized to the same rate under the 10.3~keV peak as that observed at SUL, as is done in figure~\ref{fig:ge71}, less than 10\% of the low-energy spectral excess at SUL can be assigned to partial energy depositions from $^{68}$Ge activation (both radioisotopes undergo the same decay). This $<$10\% is a conservative upper limit, given that the DAQ used in San Onofre did not feature the digitization of preamplifier traces necessary for rise time cuts (i.e., the low energy component of the $^{71}$Ge template in figure~\ref{fig:ge71} would be further reduced by those). 
      
\subsection{Environmental radon and radon daughter deposition on detector surfaces}

\begin{figure}[!htbp]
\includegraphics[height=0.22\textheight]{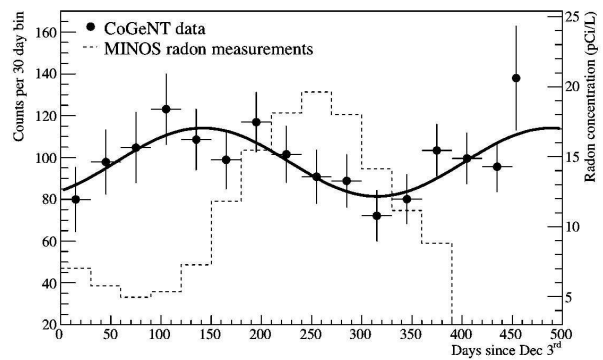}
\caption{\label{fig:radon} Counts per 30 day bins from the 0.5-3.0~keVee CoGeNT energy window (black dots) compared to the MINOS radon data at SUL (dashed), averaged over the period 2007-2011, exhibiting a peak on August 28th \protect\cite{goodm,minos}. The solid curve represents a sinusoidal fit to CoGeNT data. An analysis by the MINOS collaboration finds a three-sigma inconsistency between the phase of their measured seasonal modulation in radon concentration at SUL and CoGeNT data \protect\cite{minosmod}.}
\end{figure}

Sec.II-A describes active measures against penetration of radon into the detector's inner shielding cavity. External gamma activity from this source is efficiently blocked by the minimum of 25 cm of lead shielding around the detector (the attenuation length in lead for the highest-energy radon associated gamma emission is $\sim\!2$ cm). These measures include precautions such as automatic valving off of the evaporated nitrogen purge gas lines during replacement of the dedicated Dewar. A time analysis of the low-energy counting rate looking for signatures of radon injection (a surge followed by a decay with t$_{1/2}$=3.8 d) revealed no such instances. Radon levels at SUL are continuously measured by the MINOS experiment, showing a large seasonal variation (a factor of $\sim\!\pm$2) \cite{goodm,minos}. Figure ~\ref{fig:radon} displays a comparison between these measurements and the germanium counting rate, showing an evident lack of correlation (see also \cite{minosmod}). While we have not requested access to information regarding diurnal changes in radon level at SUL, these are commonly observed in underground sites, and seemingly absent from CoGeNT data (figure~\ref{fig:diurnal}). A modulated radon signature would appear at all energies in CoGeNT spectra, an effect not observed, due to partial energy deposition from Compton scattering of gamma rays emitted by this radioactive gas and its progeny \cite{radon}.

Additional sources of radon-related backgrounds are the delayed emissions from $^{222}$Rn daughters deposited on detector surfaces during their fabrication. The dominant low-energy radiations of concern are a beta decay with 17~keV endpoint from $^{210}$Pb, and 102~keVr lead alpha-recoils from the decay of $^{210}$Po. These radiations are known to produce a low-energy spectral rise in germanium detectors lacking sufficiently-thick protective inert surface layers \cite{edelweissthesis}. The PPC detector considered here is insulated against these over most of its surface by the $\sim$1 mm dead layer discussed in Sec.III-C. Only its intra-contact surface (3.8 cm$^{2}$) is partially sensitive to these. An inert 150 nm thick SiO$_{x}$ layer is deposited there during manufacture in order to passivate this surface, reducing leakage current across the contacts. Its thickness is almost four times the projected range of $^{210}$Po alpha recoils, effectively blocking their possible contribution. We calculate the contribution from $^{210}$Pb betas via MCNP simulation, taking as input the 90 \% C.L. upper limit to the activity of their accompanying 46.5 keV gamma emission from the spectrum in figure ~\ref{fig:data300gammas}. This upper limit translates into $<$2.8 $^{210}$Pb decays per day, conservatively assumed to take place in their entirety on the intra-contact surface. The resulting degraded beta energies reaching the surface of the active germanium are spectrally very different from the residual background observed, not exhibiting an abrupt rise near threshold, and contribute only a maximum of 5\% to the rate in the 0.5-1.5~keVee region of the irreducible spectrum in figure~\ref{fig:overlay}. We consider this upper limit to be overly conservative. 

\subsection{Backgrounds from radioactivity in cryostat materials}

Materials surrounding the CoGeNT detector are selected for their low radioactivity (Sec.II-A).  However, due to the proximity of these materials to the detector, even small activities could potentially be a background to a possible dark matter signal.  We have therefore performed simulations of these backgrounds to determine their contribution to the low-energy spectrum.  

\subsubsection{Backgrounds from OFHC Copper and PTFE}

The CoGeNT detector is contained within OFHC copper parts, etched to reduce surface contaminations (Sec.II-A).  Gamma counting of large samples of OFHC copper at Gran Sasso yield $^{238}$U and $^{232}$Th concentrations of 18~$\mu$Bq/kg and 28~$\mu$Bq/kg, respectively \cite{EHoppe}.  We have simulated the $^{238}$U and $^{232}$Th decay chains in the copper shield, including gamma emission, betas and their associated bremsstrahlung.  The simulation also includes the alpha-decays in both chains, since alpha-induced X-ray emission is potentially a background. The number of events within the 0.5-3.0~keVee region is estimated as a negligible $\sim$9 events for the entire 442~day data set in \cite{Aal11b}. A similar calculation for the 0.5~mm PTFE liner surrounding the crystal, also chemically etched, yields only 1.5 events for the same energy region and time period, using a conservative activity of 15~mBq/kg ($^{238}$U) and 7~mBq/kg ($^{232}$Th) \cite{SNOLAB}. In addition to this, we calculate an absence of measurable contribution from standard concentrations of $^{40}$K and $^{14}$C in the PTFE crystal liner ($<$85~mBq/kg and $\sim$60~Bq/kg, respectively).

\subsubsection{Backgrounds from resistors in front-end electronics}

\begin{figure}[!htbp]
\includegraphics[height=0.27\textheight]{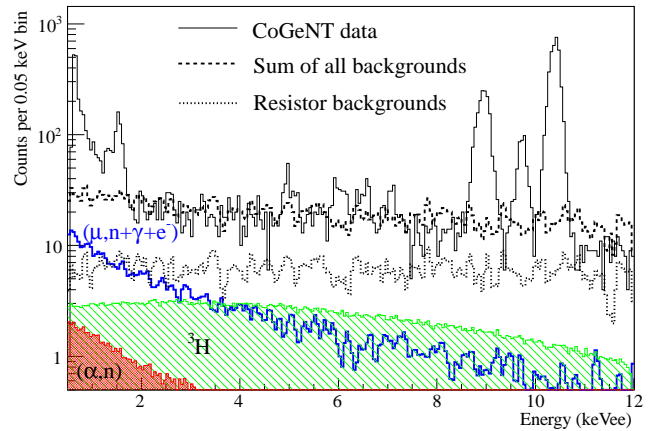}
\caption{\label{fig:resbkgs} Similar to figure \ref{fig:muoninduced}, with expanded ranges: energy spectrum of the simulated $^{238}$U,  $^{232}$Th, and $^{40}$K resistor background (dotted line) compared to CoGeNT data (solid).  In the energy range displayed the estimated resistor backgrounds are by far dominant.  The resistor background spectrum is  for  metal film resistors, the same used in the CoGeNT front end.  Also shown are other background contributions and their sum. Contributions from $^{210}$Pb bremsstrahlung and radioactivity in PTFE and OFHC cryostat parts are comparatively negligible.}
\end{figure}

The front-end FET capsule, fabricated in PTFE, contains two small resistors in close proximity (within $\sim\!2$ cm) to the germanium crystal.  Resistors are known to have relatively high levels of radioactive contaminants, and their location make them a primary candidate for the source of a large fraction of events.  Table \ref{tab:resbkgs} summarizes measured levels of $^{238}$U, $^{232}$Th, and $^{40}$K concentrations in various resistors from the ILIAS database \cite{ilias}.  The ceramic in most resistors is the largest contributor to the radioactivity.  The type of resistors used in CoGeNT are metal film on ceramic, with an approximate mass of 50~mg each.  Table \ref{tab:resbkgs} also summarizes the number of background events in the 0.5-3.0~keVee region of the 442~day data set, determined from a simulation scaled to the various activity measurements.  These range from 324$\pm$165 to 4509$\pm$352, the dominant contributions being gammas in the $^{238}$U and $^{232}$Th chains.  The spectrum of energy deposition  is shown in Figs. \ref{fig:muoninduced} and \ref{fig:resbkgs}.   These figures specifically show results for a metal film resistor, the same type of resistor  in CoGeNT, without any scaling.  Since we have not assayed the specific resistors used in CoGeNT, we cannot be certain that most of the flat background component observed in CoGeNT data is due to this source, but the agreement with this flat component of the spectrum is suggestive.  A scheme to eliminate these resistors in the C-4 design \cite{inprep} has been developed.

\begin{table*}
\caption{\label{tab:resbkgs}
Summary of measured backgrounds in various resistors from the ILIAS database \cite{ilias}, with corresponding simulated number of events in the CoGeNT 0.5-3.0~keVee region (442~day data set \cite{Aal11b}).  Uncertainties are dominated by the  activity measurement, but include the statistical uncertainty in the simulation.  The total number of expected events in this energy region  range from 324$\pm$165 to 4509$\pm$352 (see text). Two resistors at 50~mg each, as in the present CoGeNT front-end, are assumed.}
\begin{ruledtabular}
\begin{tabular}{ccccccc}
 &\multicolumn{2}{c}{$^{238}$U}&\multicolumn{2}{c}{$^{232}$Th}&\multicolumn{2}{c}{$^{40}$K}\\
Description&Rate(Bq/kg)&Events in data&Rate(Bq/kg)&Events in data&Rate(Bq/kg)&Events in data\\ \hline
carbon film resistor&4.3&269$\pm$74&12.7&687$\pm$95&21.9&16.5$\pm$4.3\\
metal film resistor 1&4.3&269$\pm$126&0.5&27$\pm$104&37.5&28.2$\pm$7.5\\
metal film resistor 2&5.1&319$\pm$99&16.1&870$\pm$125&24.7&18.6$\pm$5.7\\
ceramic core resistor&5.9&369$\pm$99&4.6&249$\pm$85&34.3&25.8$\pm$6.0\\
metal on ceramic resistor&28&1750$\pm$193&40.7&2740$\pm$294&25.7&19.4$\pm$4.7\\
\end{tabular}
\end{ruledtabular}
\end{table*}

\begin{figure}[!htbp]
\includegraphics[height=0.17\textheight]{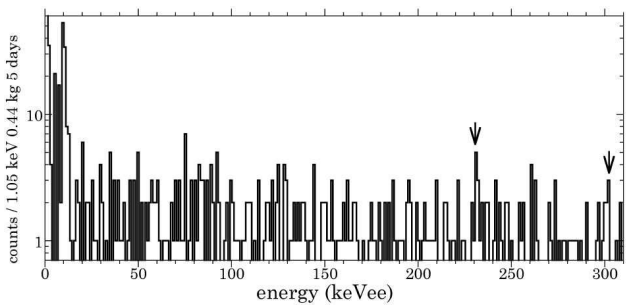}
\caption{\label{fig:data300gammas} Existing CoGeNT data in the range up to 300 keV, with possible weak $^{212}$Pb (238 keV) and $^{214}$Pb (295 keV) gamma lines indicated by arrows.  The extrapolated energy scale can only be considered approximate. The energy binning corresponds to the approximate FWHM resolution for these two lines. See text for a discussion on a possible origin for these putative lines in the front-end resistors. Notoriously absent are a  $^{210}$Pb peak at 46.5 keV and excess lead x-rays, a result of the radiopurity of the inner lead layers in the shield (Sec.II-A) and detector surfaces (Sec.V-C).}
\end{figure}

As a further consistency check we examined the existing CoGeNT data out to an energy of 300~keVee.  The statistics in this range are limited (5 days of dedicated exposure, see Sec.III-A).  Figure \ref{fig:data300gammas} shows possible 238~keV $^{212}$Pb ($^{232}$Th chain) and 295~keV ($^{238}$U chain) gamma lines. Due to their relatively-low energy, their source would be near the crystal, within the inner lead cavity. If they are considered as a measure of the $^{238}$U and $^{232}$Th chain contamination in front-end resistors, a 14$\pm$7~Bq/kg for $^{238}$U contamination and 1.6$\pm$0.7 Bq/kg for $^{232}$Th contamination is obtained for the resistors.  This activity would provide $\sim$937 events in the 0.5-3.0~keVee region, in good agreement with the measured flat component of the spectrum. The statistical evidence for these lines is however slim, and their presence is seen to be mutually exclusive when examining the uncertainties associated to the energy scale extrapolation used for this short run.
 
\subsection{Backgrounds from neutrino scattering}

While the smallness of neutrino cross-sections indicate that their contribution to the CoGeNT spectrum should be negligible, the signal from coherent neutrino-nucleus scattering \cite{freedman} from several sources (e.g. solar, atmospheric, diffuse supernova, and geo-neutrinos) would be highly concentrated at low energies. We engage here in the exercise of providing a few estimated upper limits for these contributions. Inferring from a recent analysis on solar and atmospheric neutrinos \cite{Gutlein}, a germanium
detector with 0.33 kg active mass and a $\sim$2 keV nuclear recoil
threshold (as in the present CoGeNT detector) would observe a rate of 
just $\sim$0.012 counts / year from coherent neutrino-nucleus scattering from $^{8}$B and $^{3}$He-proton fusion (HEP) solar neutrinos, the only solar sources able to produce a signal above threshold. Diffuse supernova background neutrinos and
atmospheric neutrinos might also contribute, however
their rate is reduced by factors of $>$ 10$^{4}$
 \cite{Strigari} and $>$ 10$^{5}$ \cite{Gutlein}, respectively. Geoneutrinos, having energies less than 4.5 MeV
 \cite{Monroe}, cannot produce nuclear recoil energies above the CoGeNT threshold. Each of these sources may also induce direct electron scattering. However, the
neutrino-electron scattering rate is suppressed by $\sim$ 10$^{5}$ relative to
the neutrino-nucleus coherent scattering rate \cite{Cabrera}. Therefore this other channel
cannot significantly contribute even taking into
account the factor of 32 increase in scattering targets, the absence of a quenching
factor, and the higher electron recoil energies. We notice however that interaction rates large enough to be of interest can be generated by solar neutrinos with enhanced baryonic currents \cite{maxim}. Additional mechanisms \cite{joachim} are able to generate a phenomenology involving diurnal and yearly modulations in rates.

\section{Conclusions}

CoGeNT is the first detector technology specifically designed to look for WIMP candidates in the low mass range around 10 GeV/c$^{2}$, an area of particular interest in view of existing anomalies in other dark matter experiments, recent phenomenological work in particle physics, and possible signals using indirect detection methods \cite{indirect}. However, investigation of the largely unexplored $\sim$few keV recoil energy range brings along new challenges in the understanding of low-energy backgrounds. The experience accumulated during the ongoing CoGeNT data-taking at SUL demonstrates that PPC detectors have excellent properties of long-term stability, simplicity of design, and ease of operation. This makes them highly suitable in searches for the annual modulation signature expected from dark matter particles forming a galactic halo. 

Besides their excellent energy resolution, low energy threshold and ability to reject surface backgrounds, PPCs compare well to other solid-state detectors under several criteria: a) the relative simplicity of CoGeNT's data analysis results in comparable irreducible spectra regardless of analysis pipeline, b) the response to nuclear recoils is satisfactorily understood, resulting in a reliable nuclear recoil energy scale, c) uninterrupted stable operation of PPC detectors can be expected over very long (several year) timescales.  We plan to continue improving this technology and our understanding of low-energy backgrounds within the framework of a CoGeNT expansion, the C-4 experiment \cite{inprep}.

\begin{acknowledgments}
We are indebted to Jeffrey de Jong and Alec Habig (MINOS collaboration) for sharing with us information on radon and muon rates at SUL, and to all SUL personnel for their constant support in operating the CoGeNT detector.  Work sponsored by NSF grants PHY-0653605 and PHY-1003940, The Kavli 
Foundation, and the PNNL Ultra-Sensitive Nuclear Measurement Initiative LDRD program (Information Release number PNNL-SA-90298).  N.E.F. and T.W.H. are supported by the DOE/NNSA Stewardship Science Graduate Fellowship program (grant number DE-FC52-08NA28752) and the Intelligence Community (IC) Postdoctoral Research Fellowship Program, respectively.
\end{acknowledgments}


\begin{thebibliography}{9}
\bibitem{consistency}As a consistency test, the same corrected spectrum at the bottom panel of figure~\ref{fig:steps} can be recovered by ignoring the 90\% C.L. fast pulser cut, and applying in lieu of it the overall fraction of fast events derived from fits like those in figure~\ref{fig:surface_fits}. Following a sufficiently long exposure, the distribution in rise time of surface and bulk events becomes self-evident, making possible to calculate bulk SA and surface BR for any choice of rise time cuts. This supersedes the electronic pulser method, which is limited by only providing a bulk SA, but no surface BR information.
\bibitem{gerbier1} The reduction in best-fit WIMP signal rate following the surface event correction discussed in Sec.IV-B is a factor $\sim$2 \cite{ucla,kelso}. An order of magnitude reduction is incorrectly claimed in \cite{gerbier2}.
\end{thebibliography}
\end{document}